\pdfoutput=1
\documentclass[3p]{elsarticle}

\usepackage[utf8]{inputenc}
\usepackage[british]{babel}
\usepackage{hyperref}
\usepackage{float}
\usepackage{graphicx}
\usepackage{amsfonts}
\usepackage{amsmath}
\usepackage{amsthm}
\usepackage{mathtools}
\usepackage{color}
\usepackage{caption}
\usepackage{subcaption}
\usepackage{xparse}
\usepackage{bm}
\usepackage{ifthen}
\usepackage{multirow}
\usepackage{siunitx}
\usepackage{eurosym}
\usepackage{booktabs}
\usepackage[nofillcomment,linesnumbered,vlined,boxed,commentsnumbered]{algorithm2e}
\usepackage{thmtools, thm-restate}
\usepackage{pdfpages}

\SetArgSty{textnormal}
\DeclareSIUnit{\EUR}{\text{\euro}}

\graphicspath{{./figs/bin/}}

\newtheorem{lem}{Lemma}
\newtheorem{thm}{Theorem}

\newcommand{\probabbr}{\hbox{RSECLP}}

\newcommand{\acupperbound}[1]{\mathcal{O} (#1)}

\newcommand{\nphard}{$\mathcal{NP}$-hard}

\newcommand{\defterm}[1]{\emph{#1}}

\newcommand{\integerset}{\mathbb{Z}}
\newcommand{\realset}{\mathbb{R}}
\newcommand{\nnrealset}{\realset_{\ge 0}}
\newcommand{\pintegerset}{\integerset_{>0}}
\newcommand{\nnintegerset}{\integerset_{\ge 0}}

\newcommand{\intinterval}[2]{\left[#1\,..\,#2\right]}

\newcommand{\opssetsym}{\mathcal{J}}
\newcommand{\opsset}{\opssetsym}
\newcommand{\opsym}{j}
\newcommand{\op}[1][]{\opsym_{#1}}
\newcommand{\opprime}{\opsym^{\prime}}

\newcommand{\proctimesym}{p}
\newcommand{\proctime}[1]{\proctimesym_{#1}}

\newcommand{\metintproctimesym}{\proctimesym}

\newcommand{\releasetimesym}{r}
\newcommand{\releasetime}[1]{\releasetimesym_{#1}}
\newcommand{\minreleasetime}{\min_{\op \in \opsset} \releasetime{\op}}

\newcommand{\duedatesym}{d}
\newcommand{\duedate}[1]{\duedatesym_{#1}}

\newcommand{\tardinesssym}{T}
\newcommand{\tardiness}[1]{\tardinesssym_{#1}}

\newcommand{\powersym}{P}
\newcommand{\power}[1]{\powersym_{#1}}

\DeclareDocumentCommand \schedindexing { m m } {
  \DeclareDocumentCommand #1 { o } {
    \IfNoValueTF {##1} {
      #2
    }{
      \ifthenelse{\equal{##1}{\prime} \OR \equal{##1}{*} \OR \equal{##1}{\prime\prime}}{
        {#2}^{##1}
      }{
        {#2}^{(##1)}
      }
    }
  }
}

\DeclareDocumentCommand \schedoptindexing { m m } {
  \DeclareDocumentCommand #1 { o m } {
    \IfNoValueTF {##1} {
      #2_{##2}
    }{
      \ifthenelse{\equal{##1}{\prime} \OR \equal{##1}{*} \OR \equal{##1}{\prime\prime}}{
        {#2}^{##1}_{##2}
      }{
        {#2}^{(##1)}_{##2}
      }
    }
  }
}

\newcommand{\schedsym}{\mathit{s}}
\schedindexing{\sched}{\bm{\schedsym}}
\newcommand{\startsym}{\schedsym}
\schedoptindexing{\start}{\startsym}

\newcommand{\baseschedsym}{\mathit{bs}}
\schedindexing{\basesched}{\bm{\baseschedsym}}
\newcommand{\basestartsym}{\baseschedsym}
\schedoptindexing{\basestart}{\basestartsym}
\newcommand{\baseschedmm}{\basesched[\prime]}
\newcommand{\baseschedoptmm}{\basesched[*]}
\newcommand{\basestartmm}[1]{\basestart[\prime]{#1}}
\newcommand{\basestartopt}[1]{\basestart[*]{#1}}

\newcommand{\completionsym}{C}
\newcommand{\completionord}[1]{\completionsym_{[#1]}}

\newcommand{\basestartmax}{\basestartsym^{\text{max}}}

\newcommand{\basestartbinsym}{\mathit{bs}}
\newcommand{\basestartbin}[2]{\basestartbinsym_{#1,#2}}

\newcommand{\realschedsym}{\mathit{rs}}
\schedindexing{\realsched}{\bm{\realschedsym}}

\newcommand{\realschedfnsym}{\MakeUppercase{\realschedsym}}
\DeclareDocumentCommand \realschedfn { o o } {
  \IfNoValueTF {#1} {
    \realschedfnsym
  }{
    \realschedfnsym(#1,#2)
  }
}

\newcommand{\realstartsym}{\realschedsym}
\schedoptindexing{\realstart}{\realstartsym}

\newcommand{\lateschedsym}{\mathit{ls}}
\schedindexing{\latesched}{\bm{\lateschedsym}}
\newcommand{\lateschedfnsym}{\MakeUppercase{\lateschedsym}}
\DeclareDocumentCommand \lateschedfn { o } {
  \IfNoValueTF {#1} {
    \lateschedfnsym
  }{
    \lateschedfnsym(#1)
  }
}
\newcommand{\latestartfn}[2]{\lateschedfn[#1]_{#2}}
\newcommand{\latestartsym}{\lateschedsym}
\schedoptindexing{\latestart}{\latestartsym}

\newcommand{\shiftschedsym}{\mathit{rss}}
\schedindexing{\shiftsched}{\bm{\shiftschedsym}}
\newcommand{\shiftschedfnsym}{\MakeUppercase{\shiftschedsym}}
\newcommand{\shiftstartsym}{\shiftschedsym}
\schedoptindexing{\shiftstart}{\shiftstartsym}

\newcommand{\objvalsym}{Z}
\newcommand{\objval}{\objvalsym}

\newcommand{\cutintmmsym}{I}
\newcommand{\cutintmm}[1]{\cutintmmsym^{\prime}_{#1}}

\newcommand{\mtsym}{t}
\newcommand{\mt}{\mtsym}

\newcommand{\horizonsym}{H}
\newcommand{\horizon}{\horizonsym}

\newcommand{\timeenergysym}{E^{\text{time}}}
\newcommand{\timeenergy}[1]{\timeenergysym_{#1}}

\newcommand{\lenmetintsym}{D}
\newcommand{\lenmetint}{\lenmetintsym}

\newcommand{\metintssetsym}{\Omega}
\newcommand{\metintsset}{\metintssetsym}
\newcommand{\metintsym}{\omega}
\newcommand{\metint}[1][]{\metintsym_{#1}}
\newcommand{\metintstartsym}{\tau}
\newcommand{\metintstart}[1]{\metintstartsym_{#1}}
\newcommand{\metintend}[1]{\metintstart{#1} + \lenmetint}

\newcommand{\maxenergysym}{E}
\DeclareDocumentCommand \maxenergy { o } {
  \IfNoValueTF {#1} {
    \maxenergysym^{\text{max}}
  }{
    \maxenergysym^{\text{max}}_{#1}
  }
}

\newcommand{\devsitsym}{\delta}
\newcommand{\devsym}{\devsitsym}
\newcommand{\devmax}{\devsym^{\text{max}}}
\newcommand{\devsitssetsym}{\Delta}
\newcommand{\devsitsset}{\devsitssetsym}
\DeclareDocumentCommand \devsit { o } {
  \IfNoValueTF {#1} {
    \devsitsym
  }{
    \ifthenelse{\equal{#1}{\prime}}{
      {\devsitsym}^{#1}
    }{
      {\devsitsym}^{(#1)}
    }
  }
}
\DeclareDocumentCommand \dev { o m } {
  \IfNoValueTF {#1} {
    \devsitsym_{#2}
  }{
    \ifthenelse{\equal{#1}{\prime}}{
      {\devsitsym}^{#1}_{#2}
    }{
      {\devsitsym}^{(#1)}_{#2}
    }
  }
}

\newcommand{\oppermsym}{\pi}
\newcommand{\oppermfn}[1][]{\oppermsym^{#1}}
\newcommand{\opperm}[2][]{\oppermsym^{#1} (#2)}
\newcommand{\oppermfnmm}{\oppermsym^{\prime}}
\newcommand{\oppermmm}[1]{\oppermsym^{\prime} (#1)}

\newcommand{\oppossym}{k}
\newcommand{\oppos}{\oppossym}
\newcommand{\opposmax}{\overline{\oppossym}}
\newcommand{\opposprime}{\oppossym^{\prime}}

\newcommand{\numops}{n}

\newcommand{\grahamenergylimits}{1|\releasetime{\op},\maxenergy[\metint]|\sum \tardiness{\op}}

\newcommand{\grahamwhole}{1|\releasetime{\op},\maxenergy[\metint],\devmax_{\op} = \devmax|\sum \tardiness{\op}}

\SetKwFunction{KwInitialPermutation}{GreedyInitialPermutation}
\SetKwFunction{KwComputeEarliestRobustStartTime}{ComputeEarliestRobustStartTime}
\SetKwFunction{KwComputeOptimalRobustBaselineSchedule}{ComputeOptimalRobustBaselineSchedule}
\SetKw{Fn}{Function}
\SetKw{Break}{break}
\SetKw{False}{false}
\SetKw{True}{true}
\SetKwData{KwFeasible}{OK}
\SetKwData{KwInfeasible}{INFEASIBLE\_PERMUTATION}
\SetKwBlock{BeginAlgorithm}{}{}

\newcommand{\unifdistd}[2]{\mathcal{U}\left \{#1,#2\right \}}

\newcommand{\unifdistc}[2]{\mathcal{U}\left(#1,#2\right)}

\begin{document}

\begin{frontmatter}
\includepdf[pages=1,fitpaper,noautoscale]{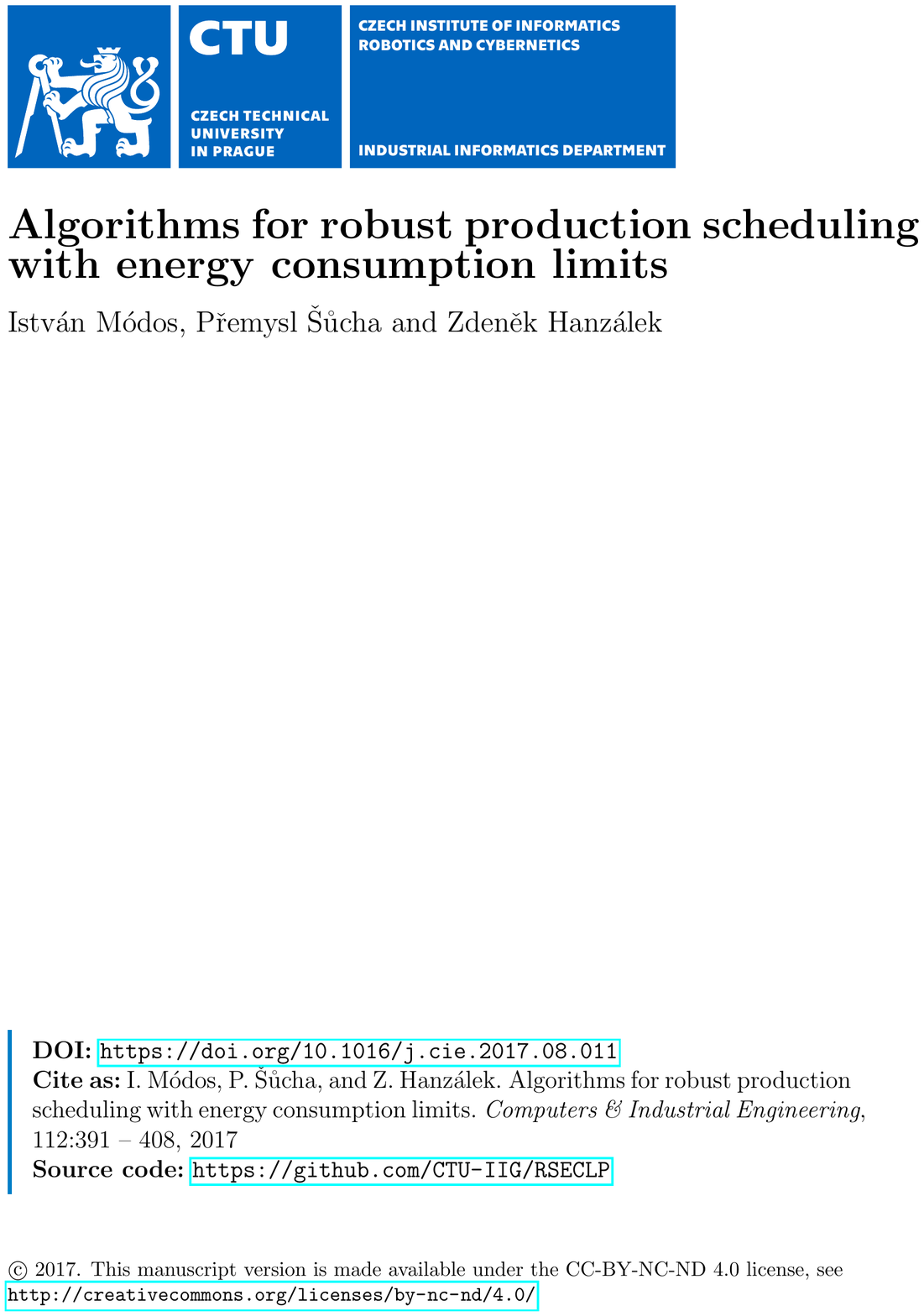}

\title{Algorithms for robust production scheduling with energy consumption limits}

 \author[dce,ciirc]{István Módos}
 \ead{modosist@fel.cvut.cz}
 \author[dce,ciirc]{Přemysl Šůcha}
 \ead{suchap@fel.cvut.cz}
 \author[dce,ciirc]{Zdeněk Hanzálek}
 \ead{hanzalek@fel.cvut.cz}

 \address[dce]{Department of Control Engineering, Faculty of Electrical Engineering, Czech Technical University, Czech Republic}
 \address[ciirc]{Czech Institute of Informatics, Robotics and Cybernetics, Czech Technical University, Czech Republic}

\begin{abstract}
  In this work, we consider a scheduling problem faced by production companies with large electricity consumption.
  Due to the contract with the electric utility, the production companies are obligated to comply with the total energy consumption limits in the specified time intervals (usually 15-minutes long); otherwise, the companies pay substantial penalty fees.
  Although it is possible to design production schedules that consider these limits as hard constraints, uncertainties occurring during the execution of the schedules are usually not taken into account.
  This may lead to situations in which the unexpected delays of the operations cause the violations of the energy consumption limits.
  Our goal is to design robust production schedules pro-actively guaranteeing that the energy consumption limits are not violated for the given set of uncertainty scenarios.
  We consider scheduling on one machine with release times of the operations and total tardiness as the objective function.

  To tackle this problem, we first propose a pseudo-polynomial algorithm for finding the optimal robust schedule for the given permutation of the operations.
  This algorithm is then utilised in three different algorithms for finding the optimal permutation: two exact (Branch-and-Bound and logic-based Benders decomposition) and one heuristic algorithm (tabu search).
  All the algorithms were experimentally evaluated on random instances with different sizes of the uncertainty scenarios set.
  Using the tabu search algorithm, we are able to solve large instances within one minute.
\end{abstract}


\end{frontmatter}


\section{Introduction}
\label{sec:intro}

In the domain of scheduling energy-demanding production, it is no longer sufficient to consider only traditional aspects such as due dates, machine capacities, tardiness, schedule length, etc.
To produce efficient schedules, the energy consumption of the operations has to be also considered~\cite{merkert2015} since significant financial savings could be achieved if the utilisation of the energy is optimised.
Although integration of the energy-awareness into production scheduling is getting more and more attention~\cite{biel2016,mansouri2016,dooren2016}, there is still a gap between industrial needs and academic research~\cite{plitsos2017}.

One of the practical problems addressed in this work is uncertainty during production in relation to the energy consumption limits.
Based on the contract with the electric utility, the companies are obligated to comply with the energy consumption limits in every 15 minutes intervals; otherwise, large penalty fees have to be paid.
However, due to the unpredictability of the operation's preparation time, it often happens that some of the operations are delayed and thus causing the violation of the contracted energy limits.
To guarantee compliance with the energy limits, reactive policies are usually employed on the shop floor.
However, using only reactive policies may lead to sub-optimal schedules or long downtimes if the schedules are not devised in a robust way (e.g.\ high and low energy-consuming operations are not alternating).

Therefore, we focus on constructing pro-active production schedules for one machine that guarantee compliance with the contracted energy consumption limits if the operations' start times are delayed within a pre-determined range; we call this a Robust Scheduling with Energy Consumptions Limits problem (\probabbr{}).

\subsection{Motivation for Robust Scheduling with Energy Consumption Limits}
The motivation for our work comes from the manufacturing and production companies with significant electricity consumption.
Specifically, we were motivated by a glass tempering process during which glass panels are heated to $620^\circ$C in a furnace.
In the considered scheduling problem, the furnace is a resource, and the heating of the glass panels represent the operations to be scheduled.
Due to technological requirements, heating of the glass panels cannot be interrupted (i.e.\ preemption is not allowed).
Although the production process also contains pre-processing and post-processing stages, we consider only scheduling of the heating stage because it is the most energy-demanding one.
However, the pre-processing and the post-processing production stages are not completely ignored since they appear as release times and due dates of the operations, respectively.
To ensure the smoothness of the production, it is reasonable to minimise the total tardiness.

\begin{figure}[H]
  \centering
    \includegraphics[width=0.38\textwidth]{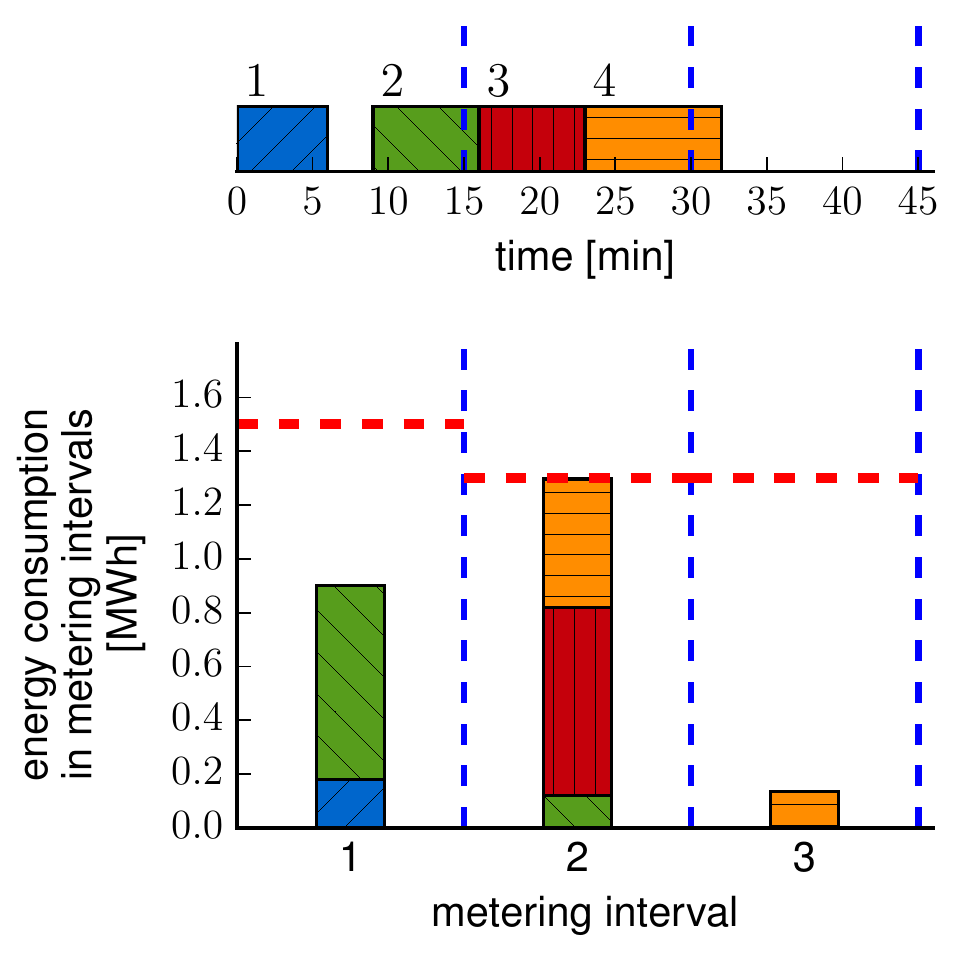}
    \caption{A baseline schedule of four operations $\{1,2,3,4\}$ on one machine which satisfies the energy consumption limits in every metering interval $\{1,2,3\}$.
The energy consumption limits are denoted by the dashed horizontal red lines.
    }
\label{fig:intro/feasible-baseline}
\end{figure}

According to the negotiated contract with the electric utility, the companies are obligated to keep their power demand below a contracted \textit{maximum power demand}.
Otherwise, the companies pay substantial penalty fees; in the Czech Republic, the penalty is regulated, and it is approximately \SI{10000}{\EUR} per consumed \SI{}{\mega\watt} over the maximum power demand~\cite{eru102016}.
The measurement of the demand is taken in every 15 minutes \defterm{metering interval} of a day, and it is measured as an average power demand during the corresponding metering interval.
Since the consumed energy can be computed as a product of the average power demand and the length of the metering interval, an equivalent formulation is that in every metering interval the total energy consumption cannot exceed the \defterm{maximum energy consumption limit}.
By considering a proper order of the energy-demanding operations or inserting short idle times, it is possible to design production schedules that do not violate these energy limits, e.g.\ see Fig.~\ref{fig:intro/feasible-baseline}.

\begin{figure}[H]
  \centering
  \begin{subfigure}[t]{0.38\textwidth}
    \includegraphics[width=\textwidth]{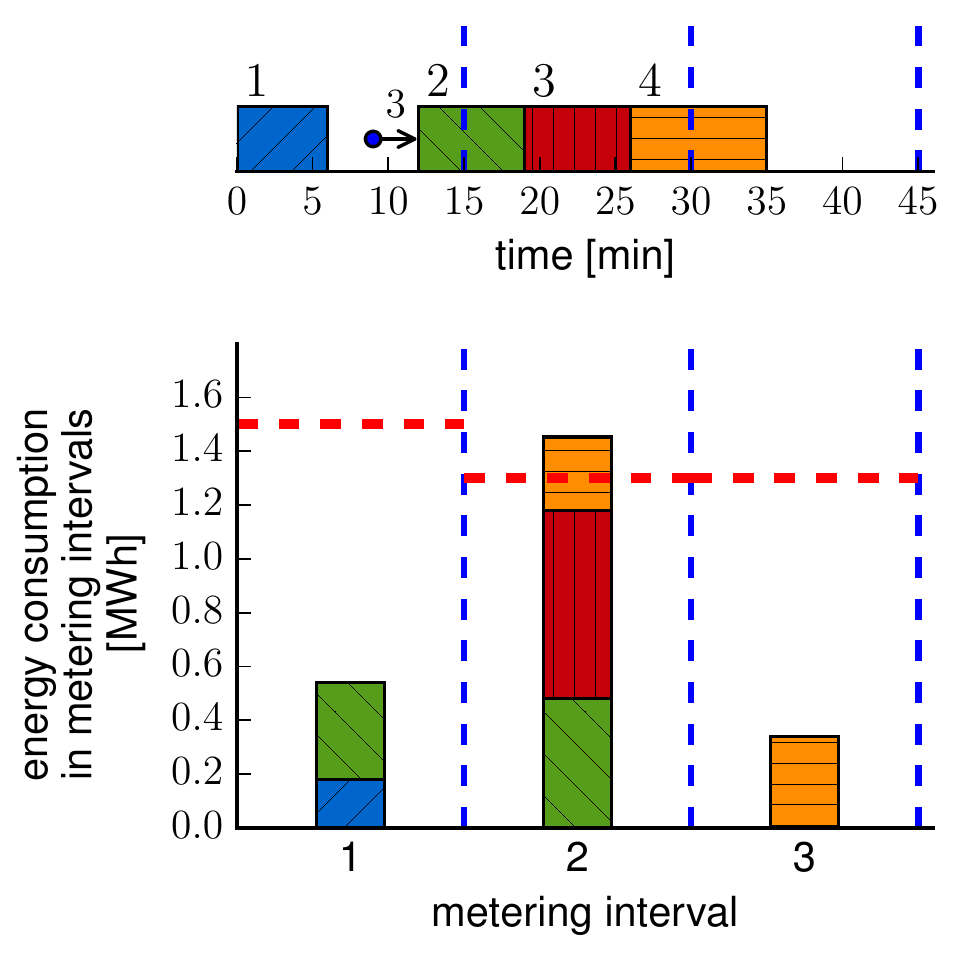}
    \caption{An example of a realised schedule in which operation 2 is delayed by 3 minutes from its baseline start time (see Fig.~\ref{fig:intro/feasible-baseline}) thus increasing the total energy consumption in metering interval 2 above the energy limit.}
\label{fig:intro/realised-violating}
  \end{subfigure}
  \quad
  \quad
  \quad
  \quad
  \quad
  \quad
  \begin{subfigure}[t]{0.38\textwidth}
    \includegraphics[width=\textwidth]{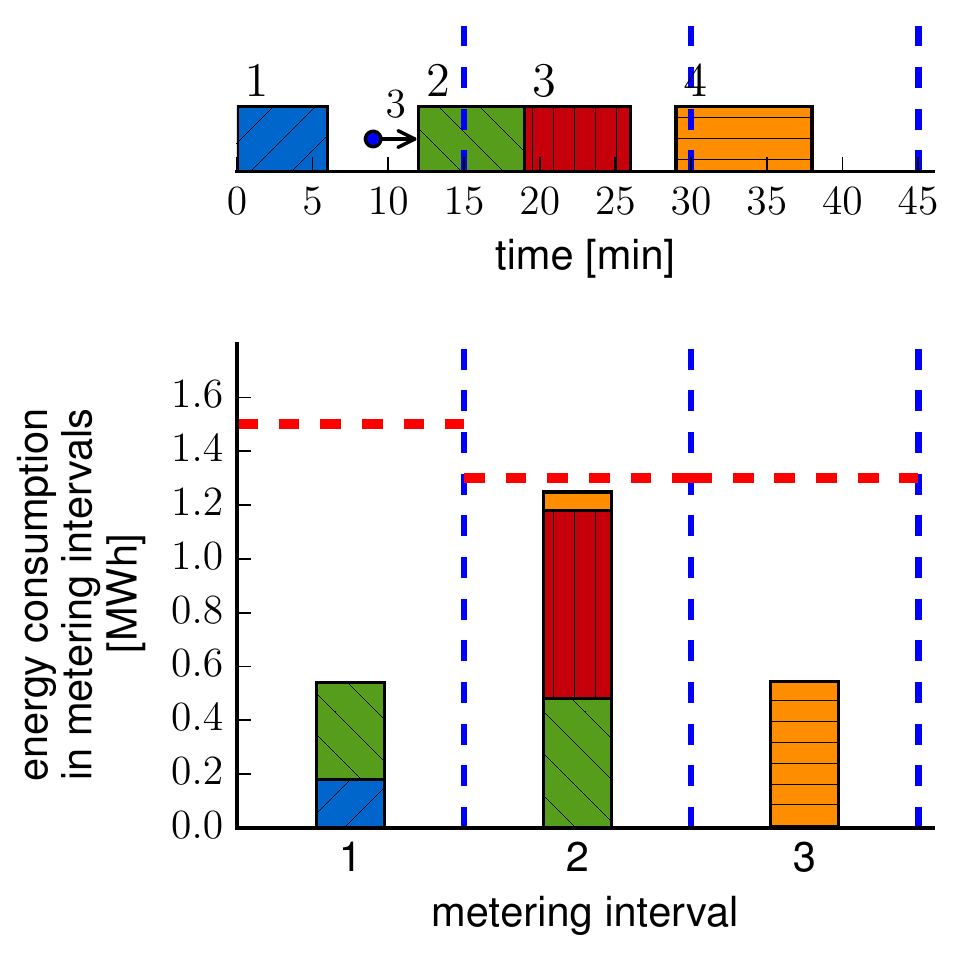}
    \caption{An application of a reactive policy on schedule from Fig.~\ref{fig:intro/realised-violating}. Operation 4 is forcibly delayed so that the energy consumption limit is not violated in metering interval 2.}
\label{fig:intro/reactive-realised-moved}
  \end{subfigure}
  \caption{
    Handling of operations' delays by a reactive policy.
  }
\end{figure}

However, in reality, unexpected events can cause delays of the operations' start times.
In the glass production example, the responsible worker has to carefully put the glass panels on the furnace conveyor, mark the panels and set the furnace parameters before the glass panels are heated.
This preparation process may take minutes, and due to various reasons (inexperienced seasonal workers, delays in the preceding production stages, etc.), it might happen that the heating of a glass panel starts later than expected.
We call the delayed start time a \defterm{realised start time}, whereas the initial non-delayed start time is referred to as a \defterm{baseline start time}.
The issue is that delaying an operation may cause an increase in the energy consumption in some metering interval above the energy limit if the energy demanding operations are started consecutively in the baseline schedule, e.g.\ see Fig.~\ref{fig:intro/realised-violating}.
In such a situation, the company pays the penalty fee even though the baseline schedule (see Fig.~\ref{fig:intro/feasible-baseline}) does not violate the energy limits.
Therefore, to design robust baseline schedules, these uncertainties have to be considered so that the energy consumption limits are not violated and the penalty fees are avoided.

\begin{figure}[H]
  \centering
  \begin{subfigure}[t]{0.38\textwidth}
    \includegraphics[width=\textwidth]{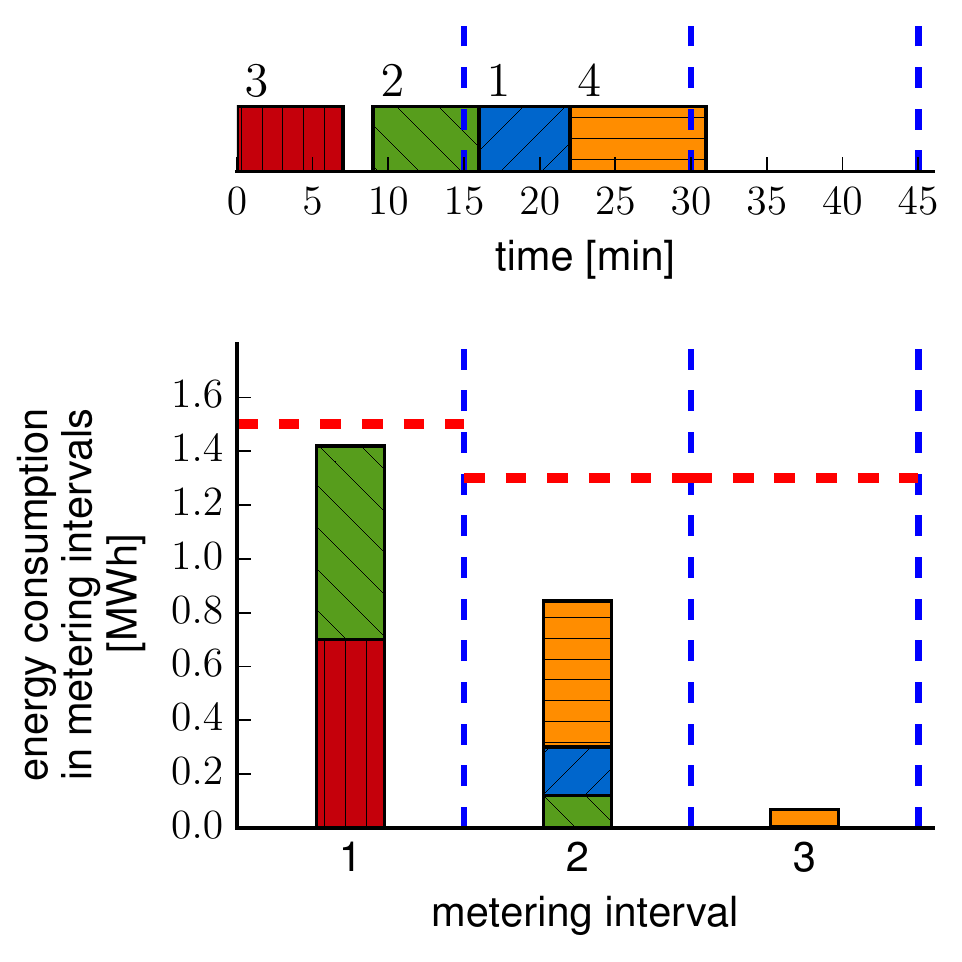}
    \caption{A robust baseline schedule of four operations $\{1,2,3,4\}$ on one machine which satisfies the energy consumption limits in every metering interval $\{1,2,3\}$.
    }
\label{fig:intro/robust-baseline}
  \end{subfigure}
  \quad
  \quad
  \quad
  \quad
  \quad
  \quad
  \begin{subfigure}[t]{0.38\textwidth}
    \includegraphics[width=\textwidth]{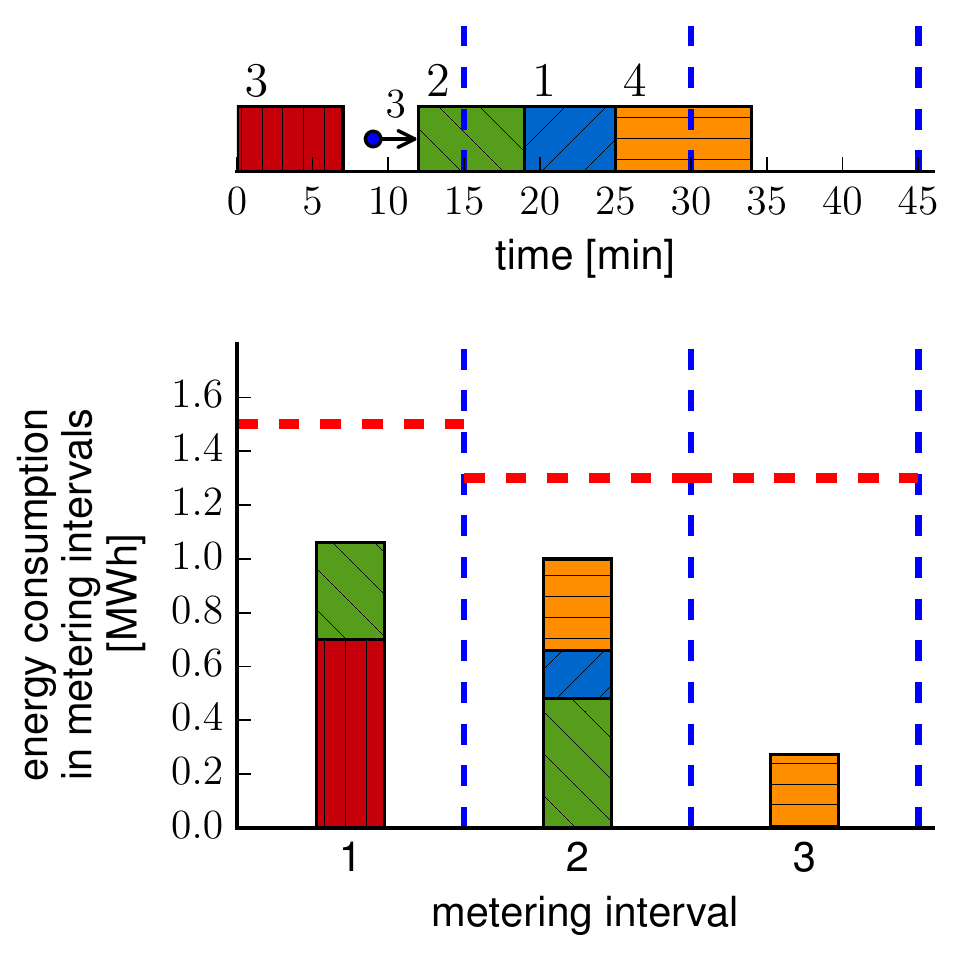}
    \caption{An example of a realised schedule in which operation 2 is delayed by 3 minutes from its baseline start time (see Fig.~\ref{fig:intro/robust-baseline}).
    Even though the total energy consumption increased in metering interval 2, the energy limit is not violated.}
\label{fig:intro/robust-realised}
  \end{subfigure}
  \caption{
    Handling of operations' delays by a pro-active scheduling.
  }
\end{figure}

One possible approach to tackling these uncertainties is to employ \defterm{reactive scheduling policies}, i.e.\ when the total energy consumption approaches the energy limit, the remaining unfinished operations are delayed until the start of the next metering interval.
For example, in Fig.~\ref{fig:intro/reactive-realised-moved} the start time of operation 4 is forcibly delayed by a worker responsible for monitoring the production process.
However, relying only on the reactive policies may cause long downtimes in the production if the order of the operations is not chosen reasonably in a baseline schedule.
A more viable approach is to combine the reactive policies with a \defterm{pro-active scheduling}, i.e.\ the baseline schedule is designed in such a way that the hazardous situations are avoided if the deviations of the operations are reasonably small.
For example, if the order of the operations from Fig.~\ref{fig:intro/feasible-baseline} would be $(3,2,1,4)$, as illustrated in Fig.~\ref{fig:intro/robust-baseline}, then even if operation 2 is delayed by 3 minutes the energy consumption limits are not violated (see Fig.~\ref{fig:intro/robust-realised}).
However, longer production delays (e.g.\ furnace breakdown) are still handled by reactive policies or by a complete rescheduling.

\subsection{Related Work}
The related work to the \probabbr{} can be categorised into two main groups: scheduling with energy constraints and robust scheduling.

The problem of maximum power demand was studied in~\cite{fangkan2013,bruzzone2012}, although the models presented in these works do not consider 15-minutes intervals but rather complying with the maximum power demand at every time instant.
Another related problem to the maximum energy consumption limits is the problem of \textit{electrical load tracking}~\cite{nolde2010,hait2011b,hait2011,hadera2015}, where the objective is to minimise the absolute difference between the actual and pre-agreed energy consumption over all metering intervals w.r.t. \hbox{penalty-free} deviation range.
Contrary to the \probabbr{}, both over-consumption and under-consumption of the energy are penalised in the load tracking problem.

Robust scheduling is a well-studied problem in the domain of \textit{resource constrained project scheduling}~\cite{herroelen2007,lambrechts2007}.
The robustness is obtained either by a robust resource allocation or inserting time buffers between activities.
In the domain of the resource constrained project scheduling, the closest problem to the \probabbr{} is presented in~\cite{policella2007}.
The goal of this work is to find a partial-ordering of the activities so that if the activities are arbitrarily delayed (w.r.t.\ to the ordering), the total demand of the resources in every time instant is below the respective capacities.
The difference from our problem is that we limit the integral of the operations' demands w.r.t.\ the intersection length of the operations with the metering intervals.

A particular interest for us is the modelling using \defterm{uncertainty scenarios}~\cite{coban2016,cheref2016}, which are used when the probability distribution of uncertain events is either not known or is uniform.
An uncertainty scenario is one realisation of uncertain events, e.g.\ a time occurrence of a machine breakdown.
In general, the objective of the scheduling with uncertainty scenarios is to mitigate the worst-case execution over all uncertainty scenarios.

To the best of our knowledge, only few works deal with both robust scheduling and energy limits.
One of such works is~\cite{nagasawa2015}, where the goal is to reduce the peak power consumption of flow shop schedules under uncertain processing times of the operations. 
The method proposed by the authors inserts idle times into the schedule to reduce the expected peak power demand.
The time points for idle times are computed by evaluating all the possible schedules originating from the set of possible scenarios and, therefore, the running time of the algorithm can increase significantly with the size of this set.

\subsection{Contribution and Outline}
Our work addresses an issue of integrating energy-awareness with robust production scheduling; we argue that this is necessary so that scheduling algorithms can be used in practice.
As seen from the related literature section, no work considers this problem in its entirety.

The main contributions of this paper are:
\begin{enumerate}
  \item A pseudo-polynomial algorithm that finds, for a given permutation of the operations, a \emph{robust} and \emph{optimal schedule} w.r.t.\ the total tardiness objective function.
    The algorithm can be incorporated into a wide variety of methods for solving the \probabbr{} that are based on searching the space of the permutations of the operations.
    To demonstrate the universality of the algorithm we employ it in two exact and one heuristic algorithms for solving the \probabbr{} (see Contributions 2 a 3).

    The complexity of the algorithm is pseudo-polynomial since it is linear in the maximum deviation of the operations (see Section~\ref{sec:optsched/complete}).

  \item Two exact algorithms (see Section~\ref{sec:exact}) for solving the \probabbr{}: (i) a Branch-and-Bound algorithm and (ii) a logic-based Benders decomposition algorithm with no-good cuts based on the optimal robust schedules.

  \item A tabu search heuristic (see Section~\ref{sec:heur}) for solving the \probabbr{}.
    The experiments (see Section~\ref{sec:exp}) show that the heuristic can solve instances with 100 operations within a minute.
\end{enumerate}

We partly studied this problem in the conference paper~\cite{modos2016a}, where we introduced the \probabbr{} and presented (i) a decomposition algorithm with simple cuts and (ii) a procedure for deciding whether the given schedule is robust or not.
Except for the master problem used in the logic-bases Benders decomposition algorithm (see Section~\ref{sec:exact/master}), all the proposed contributions of this paper, as specified above, are novel.

The paper is organised as follows.
Section~\ref{sec:prob} states the problem in a formal way.
Section~\ref{sec:optsched} describes how the optimal robust baseline schedule is constructed for the given permutation of the operations.
The next Section~\ref{sec:exact} is concerned with the exact algorithms, and Section~\ref{sec:heur} describes the heuristic algorithm.
In Section~\ref{sec:exp} the proposed algorithms are experimentally evaluated.
Finally, the last section concludes the paper.


\section{Problem Statement}
\label{sec:prob}
The production scheduling problem outlined above is formally defined in this section.
First, the scheduling problem considering only the energy consumption limits without robustness is described in Section~\ref{sec:prob/nonrobust}.
This scheduling problem is then extended in Section~\ref{sec:prob/robust} with the deviations of the start times.

In the rest of the text, we use the notation $\intinterval{a}{b} = [a, b] \cap \integerset$ to denote the integer interval for given $a, b \in \integerset$.
Moreover, the length of intersection of two intervals will be denoted as \hbox{$\text{lenint}([a_1,b_1],[a_2,b_2]) = \max(0, \min(b_1, b_2)-\max(a_1, a_2))$}.

\subsection{Non-robust Scheduling with Energy Consumption Limits Problem}
\label{sec:prob/nonrobust}
Let \mbox{$\opsset = \intinterval{1}{\numops}$} be a \defterm{set of operations} that have to be scheduled on a single machine without preemption.
For each \defterm{operation} \mbox{$\op \in \opsset$} we define \defterm{release time} $\releasetime{\op} \in \nnintegerset$, \defterm{processing time} $\proctime{\op} \in \pintegerset$ and \defterm{due date} $\duedate{\op} \in \nnintegerset$.
Moreover, for each operation $\op \in \opsset$ we also define $\power{\op} \in \nnrealset$ representing the \defterm{power consumption} of the machine when processing operation $\op$, i.e.\ it is the constant rate at which the energy is consumed in every time instant.
Therefore, the total consumed energy by each operation $\op$ is $\proctime{\op} \cdot \power{\op}$.

The operations have to be scheduled on a single machine within \defterm{scheduling horizon} $\horizon \in \nnintegerset$, i.e.~the operations must complete at most at time $\horizon$.
The scheduling horizon is divided into a \defterm{set of metering intervals} $\metintsset = \intinterval{1}{\frac{\horizon}{\lenmetint}}$ with equal \defterm{length} of $\lenmetint \in \pintegerset$ (it is assumed that $\horizon$ is a multiple of $\lenmetint$).
For each metering interval $\metint \in \metintsset$, an \defterm{energy consumption limit} is denoted as $\maxenergy[\metint]$, which represents the upper bound on the total energy consumption of the operations in metering interval $\metint$.
Moreover, let us denote a \defterm{start} of interval $\metint$ as $\metintstart{\metint} = (\metint - 1) \cdot \lenmetint$ and its \defterm{end} as $\metintstart{\metint} + \lenmetint$.

\defterm{Baseline schedule} $\basesched \in \nnintegerset^{\numops}$ is a vector, where each element $\basestart{\op}$ represents the \textit{baseline start time} of operation $\op \in \opsset$ and the operations are not overlapping in $\basesched$.
If operation $\op$ starts at time $\basestart{\op}$, then a \defterm{tardiness} of operation $\op$ is defined as \hbox{$\tardiness{\op} = \max(0, (\basestart{\op} + \proctime{\op}) - \duedate{\op})$}.
Moreover, the \defterm{intersection length} between metering interval $\metint \in \metintsset$ and operation $\op \in \opsset$ starting at time $\mt$ is denoted as \hbox{$\metintproctimesym(\metint,\op,\mt) = \text{lenint}([\metintstart{\metint},\metintend{\metint}],[\mt,\mt + \proctime{\op}])$}.

The goal of this scheduling problem is to find baseline schedule $\basesched$ such that
\begin{align}
  \min \quad&\sum_{\op \in \opsset} \tardiness{\op}  \\
  \label{eq:prob/nonrobust/release-time}
  \mbox{s.t.}\quad&\releasetime{\op} \le \basestart{\op}, \; \op \in \opsset \\
  \label{eq:prob/nonrobust/max-start}
  &\basestart{\op} + \proctime{\op} \le \horizon, \; \op \in \opsset \\
  \label{eq:prob/nonrobust/limit-feasible}
  &\sum_{\op \in \opsset} \metintproctimesym(\metint,\op,\basestart{\op}) \cdot \power{\op} \le \maxenergy[\metint], \; \metint \in \metintsset
\end{align}
where Constraint~\eqref{eq:prob/nonrobust/limit-feasible} enforces that the energy consumption limit is not violated in any metering interval.
We describe this problem in Graham's notation~\cite{graham1979} as $\grahamenergylimits$.

\subsection{Robust Scheduling with Energy Consumption Limits Problem (\probabbr{})}
\label{sec:prob/robust}
Since unexpected events may occur during the execution of a baseline schedule, the actual start times of the operations may be delayed from their baseline start times; we call the carried out schedule \defterm{a realised schedule}.
The goal of the robust scheduling is to guarantee that the energy consumption limits are not violated in any possible realised schedule.

To formally define the realised schedules, the notion scenarios and the maximum deviation has to be introduced.
Let $\devmax \in \nnintegerset$ be a \defterm{maximum deviation} of any operation.
Then \hbox{$\devsitsset = {\intinterval{0}{\devmax}}^{\numops}$} is a \defterm{set of all scenarios} such that \defterm{scenario} $\devsit \in \devsitsset$ is a vector where each element $\dev{\op}$ represents the \defterm{deviation} of operation $\op \in \opsset$.
The maximum deviation is a user parameter which can be set according to the required range of the covered realised schedules.

Let $\oppermfn: \intinterval{1}{\numops} \rightarrow \opsset$ be a bijective function representing a \defterm{permutation of the operations}.
Operation on $\oppos$-th \defterm{position} in permutation $\oppermfn$ is denoted as $\opperm{\oppos}$.
We will say that $\oppermfn$ is the \defterm{corresponding} permutation of $\basesched$ if the order of operations in $\basesched$ is the same as in $\oppermfn$.

From baseline schedule $\basesched$, its corresponding permutation $\oppermfn$ and arbitrary scenario $\devsit \in \devsitsset$, one can derive \defterm{a realised schedule} using recursive vector function $\realschedfn$ as
\begin{equation}
  \label{eq:prob/realsched}
  \begin{split}
    \realschedfn[\basesched][\devsit]_{\opperm{\oppos}} =
    \begin{dcases}
      \basestart{\opperm{1}} + \dev{\opperm{1}} & \oppos = 1 \\
      \max(\basestart{\opperm{\oppos}}, \realschedfn[\basesched][\devsit]_{\opperm{\oppos - 1}} + \proctime{\opperm{\oppos - 1}})+ \dev{\opperm{\oppos}} & \text{otherwise.}
    \end{dcases}
  \end{split}
\end{equation}
Please notice that the definition of the realised schedules implies $\dev{\opperm{\oppos}} \le\realschedfn[\basesched][\devsit]_{\opperm{\oppos}} - \basestart{\opperm{\oppos}}$; to make the distinction clear, the value of $\realschedfn[\basesched][\devsit]_{\opperm{\oppos}} - \basestart{\opperm{\oppos}}$ is called a \defterm{delay}.
To summarise, deviation $\dev{\op}$ of operation $\op$ is independent of the deviations of the other operations, whereas the delay is not.

To ensure that no operation completes outside of horizon $\horizon$ even if all operations are delayed, a \defterm{maximum start time} is defined as \hbox{$\basestartmax = \horizon - (\numops \cdot \devmax + \max_{\op \in \opsset} \proctime{\op})$} where value $\numops \cdot \devmax$ represents the maximum possible delay of any operation in any realised schedule from its baseline start time.
Although $\basestartmax$ is a pessimistic bound w.r.t.\ horizon $\horizon$, it is simple to compute and can be incorporated into algorithms requiring a solution-independent upper bound on the baseline start times.
Moreover, since $\basestartmax$ increases with the scheduling horizon (which is a user-provided parameter), setting the horizon large enough will loosen $\basestartmax$.

The goal of the \probabbr{} is to find baseline schedule $\basesched$ such that
\begin{align}
  \min \quad &\sum_{\op \in \opsset} \tardiness{\op}  \\
  \label{eq:prob/robust/release-time}
  \mbox{s.t.}\quad&\releasetime{\op} \le \basestart{\op}, \; \op \in \opsset \\
  \label{eq:prob/robust/max-start}
  &\basestart{\op} \le \basestartmax, \; \op \in \opsset \\
  \label{eq:prob/robust/robust}
  &\sum_{\op \in \opsset} \metintproctimesym(\metint,\op,\realschedfn[\basesched][\devsit]_{\op}) \cdot \power{\op} \le \maxenergy[\metint], \; \metint \in \metintsset,\devsit \in \devsitsset
\end{align}
where Constraint~\eqref{eq:prob/robust/robust} enforces that energy consumption limit is not violated in any metering interval in any realised schedule (or, equivalently, in no scenario).
Notice that Constraint~\eqref{eq:prob/robust/robust} reduces to Constraint~\eqref{eq:prob/nonrobust/limit-feasible} when $\devsit = (0, 0, \dots, 0)$.
A baseline schedule which does not violate Constraint~\eqref{eq:prob/robust/robust} is called \defterm{a robust baseline schedule}.
We describe the \probabbr{} in Graham's notation as $\grahamwhole$.

\begin{figure}[H]
  \centering
  \includegraphics[scale=0.7]{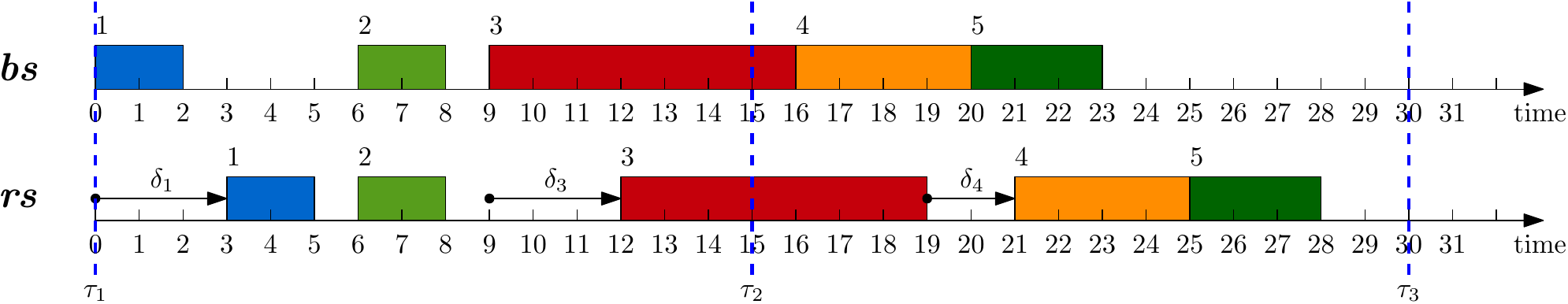}
  \caption{A visualisation of the example.}
\label{fig:prob/example}
\end{figure}

The \probabbr{} is illustrated on the following simple example with 5 operations $\opsset = \{1,2,3, 4, 5\}$.
Let $\lenmetint = 15$, $\metintsset = \intinterval{1}{5}$, $\devmax = 3$ and $\maxenergy[\metint] = 1200$, where $\metint \in \metintsset$.
The parameters of the operations are provided in \hbox{Tab.~\ref{tb:prob/example-operations-values}}.
One particular baseline schedule $\basesched$ is shown in \hbox{Tab.~\ref{tb:prob/example-baseline}} and realised schedule $\realsched = \realschedfn[\basesched][\devsit]$ for scenario \hbox{$\devsit = (3, 0, 3, 2, 0)$} is provided in \hbox{Tab.~\ref{tb:prob/example-realised}}.
The visualisation of the baseline and the realised start times is in Fig.~\ref{fig:prob/example}.
Notice that $\realstart{4} - \basestart{4} = 5 > \devmax$ due to the delay of operation 3.
Moreover, $\realstart{5} - \basestart{5} > 0$ even though $\dev{5} = 0$; this is due to the delays of the preceding operations.
The total tardiness in the baseline schedule equals to $4$ and the total energy consumption in the realised schedule in metering intervals 1 and 2 are $2\cdot 50 + 2\cdot 70 + 3\cdot 150 + 0 \cdot 120 + 0 \cdot 30=690$ and $0 \cdot 50 + 0 \cdot 70 + 4\cdot 150+ 4\cdot 120 + 3\cdot 30=1170$, respectively.
Therefore, the energy consumption limits are not violated in $\realsched$ (the energy consumption in every other metering interval is 0).

\begin{table}[H]
  \centering
  \captionsetup{labelsep=newline,justification=centering}
  \begin{minipage}[t]{.25\linewidth}
    \centering
    \caption{Parameters of the operations.}
\label{tb:prob/example-operations-values}
    \begin{tabular}{lcccc}
      \toprule
      $\op$&$\releasetime{\op}$&$\duedate{\op}$&$\proctime{\op}$&$\power{\op}$\\
      \midrule
      1&0&5&2&50\\
      2&6&10&2&70\\
      3&8&15&7&150\\
      4&10&17&4&120\\
      5&18&30&3&30\\
      \bottomrule
    \end{tabular}
  \end{minipage}
  \begin{minipage}[t]{.25\linewidth}
    \centering
    \caption{Baseline schedule $\basesched$.}
\label{tb:prob/example-baseline}
    \begin{tabular}{lc}
      \toprule
      $\op$&$\basestart{\op}$\\
      \midrule
      1&0\\
      2&6\\
      3&9\\
      4&16\\
      5&20\\
      \bottomrule
    \end{tabular}
  \end{minipage}
  \begin{minipage}[t]{.4\linewidth}
    \centering
    \caption{Realised schedule $\realsched$ for $\devsit = (3,0,3,2,0)$.}
\label{tb:prob/example-realised}
    \begin{tabular}{lccc}
      \toprule
      $\op$&$\realstart{\op}$&$\metintproctimesym(1,\op,\realstart{\op})\cdot \power{\op}$&$\metintproctimesym(2,\op,\realstart{\op})\cdot\power{\op}$\\
      \midrule
      1&3&$2 \cdot 50$&$0 \cdot 50$\\
      2&6&$2\cdot 70$&$0 \cdot 70$\\
      3&12&$3\cdot 150$&$4\cdot 150$\\
      4&21&$0\cdot 120$&$4\cdot 120$\\
      5&25&$0\cdot 30$&$3\cdot 30$\\
      \bottomrule
    \end{tabular}
  \end{minipage}
\end{table}

The \nphard{}ness of the \probabbr{} was shown in~\cite{modos2016a}, which follows from the underlying problem $1||\sum \tardiness{\op}$.
Moreover, even checking the robustness of a given baseline schedule is not an easy problem at first sight since a na{\"\i}ve algorithm would check all ${(\devmax + 1)}^{\numops}$ scenarios.


\section{Algorithm for Finding Optimal Robust Schedule for Fixed Permutation}
\label{sec:optsched}

As it has been shown at the end of the previous section, the \probabbr{} is a difficult combinatorial problem.
To solve it, we first focus on a simpler, related problem: is it possible to find an optimal robust schedule (i.e.\ satisfying Constraints~\eqref{eq:prob/robust/release-time}-\eqref{eq:prob/robust/robust}) for a fixed permutation $\oppermfn$ of the operations quickly?
If the answer is ``yes'', then the \probabbr{} can be solved by a natural decomposition into two parts: (i) search the space of the permutations and (ii) for the given permutation, find the optimal robust schedule.
  
In this section, we introduce an algorithm with pseudo-polynomial complexity of $\acupperbound{\numops^3 \cdot \devmax + \numops \cdot |\metintsset|}$ that creates an optimal robust schedule from the given permutation $\oppermfn$ of the operations.
This algorithm is the cornerstone of the exact and the heuristic approaches described in Section~\ref{sec:exact} and Section~\ref{sec:heur}, respectively.
To better explain the main concepts behind the algorithm, the description is split into several subsections.

\subsection{Latest start time and right-shift schedules}
Before we explain the algorithm, the notions of \defterm{latest start times} and \defterm{right-shift schedules} have to be defined.
Both definitions assume some fixed baseline schedule $\basesched$ and its corresponding permutation $\oppermfn$.

\defterm{Latest start time schedule} $\latesched$ is defined using vector function $\lateschedfn$ as
\begin{equation}
  \latestartfn{\basesched}{\opperm{\oppos}} = \realschedfn[\basesched][(\devmax, \devmax, \dots, \devmax)]_{\opperm{\oppos}},\; \oppos \in \intinterval{1}{\numops}
\end{equation}
It represents the maximum possible starting time over all realised schedules for fixed baseline schedule $\basesched$.

Let $\opperm{\opposprime}$ be some operation and $\mt \in \intinterval{\basestart{\opperm{\opposprime}}}{\latestartfn{\basesched}{\opperm{\opposprime}}}$ its arbitrary realised start time.
Then \defterm{right-shift schedule} $\shiftsched$ is defined using recursive vector function $\shiftschedfnsym$ as
\begin{equation}
  \shiftschedfnsym(\basesched, \opposprime, \mt)_{\opperm{\oppos}} =
  \begin{dcases}
    \mt & \oppos = \opposprime \\
    \min(\latestartfn{\basesched}{\opperm{\oppos}}, \shiftschedfnsym(\basesched, \opposprime, \mt)_{\opperm{\oppos + 1}} - \proctime{\opperm{\oppos}}) & \oppos \in \intinterval{1}{\opposprime - 1}
  \end{dcases}
\end{equation}
Informally, a right-shift schedule is obtained from $\basesched$ by fixing the start time of operation $\opperm{\opposprime}$ to $\mt$ and shifting all the operations on positions $\oppos < \opposprime$ to the right as much as possible while respecting the latest start times and the no-overlap constraint.
Notice that a right-shift schedule defines starting times only for the operations on positions $\intinterval{1}{\opposprime}$.

An important property of the right-shift schedules is that they are also realised schedules, i.e.\ for each right-shift schedule $\shiftsched$ there exists scenario $\devsit \in \devsitsset$ whose corresponding realised schedule $\realsched$ is the same as $\shiftsched$ (see Lemma~\ref{lem:optsched/shiftdev} in Appendix).

\subsection{Earliest robust baseline schedule}
The algorithm for finding the optimal robust schedule is based on the iterative computation of the \defterm{earliest robust baseline start time} for each operation in the order given by $\oppermfn$.
Robust baseline start time of $\opperm{\opposmax}$ is a baseline start time such that there is no realised schedule of operations $\opperm{1},\opperm{2},\dots,\opperm{\opposmax}$ in which some energy consumption limit is violated.
More formally, $\basestart{\opperm{\opposmax}}$ is \defterm{robust relative to} baseline start times $\basestart{\opperm{1}},\basestart{\opperm{2}},\dots,\basestart{\opperm{\opposmax - 1}}$ if
\begin{equation}
  \sum_{\oppos=1}^{\opposmax} \metintproctimesym(\metint,\opperm{\oppos},\realschedfn[\basesched][\devsit]_{\opperm{\oppos}}) \cdot \power{\opperm{\oppos}} \le \maxenergy[\metint], \quad \metint \in \metintsset,\devsit \in \devsitsset
\label{eq:optsched/robust-start-time}
\end{equation}
The \defterm{earliest robust baseline start time} is simply a robust baseline start time that is the smallest possible relative to the baseline start times of the preceding operations.
It can be proven (see Theorem~\ref{thm:app/earliest-schedule-is-robust-optimal} in Appendix) that baseline schedule $\basesched$ is robust and optimal w.r.t.\ the total tardiness objective function if every operation starts at its earliest robust time in $\basesched$.
Since the earliest robust start time of $\opperm{\opposmax}$ depends only on the baseline start times of $\opperm{1},\opperm{2},\dots,\opperm{\opposmax-1}$, the earliest robust start times can be computed one-by-one according to the ascending order of the positions in the given permutation $\oppermfn$, see Algorithm~\ref{alg:optsched/schedule}.
\begin{algorithm}[H]
  \caption{Optimal Robust Baseline Schedule for Fixed Permutation}
\label{alg:optsched/schedule}
  \Fn{}
  \KwComputeOptimalRobustBaselineSchedule{$\oppermfn$}
  \BeginAlgorithm{
    \DontPrintSemicolon{}
    $\basesched \gets (\infty, \infty, \dots, \infty)$\;
    $\latesched \gets (\infty, \infty, \dots, \infty)$\;
    \ForEach{$\opposmax = 1,\dots,\numops$}{
      \tcc{Computation of the earliest robust baseline start time of $\opperm{\opposmax}$.}
      \If{$\KwComputeEarliestRobustStartTime(\oppermfn, \opposmax, \basesched, \latesched) = \KwInfeasible$}{
        \Return{$\KwInfeasible,\emptyset$}\;
      }
    }
    \Return{$\KwFeasible,\basesched$}\;
  }
\end{algorithm}

Algorithm~\ref{alg:optsched/schedule} terminates either with computing the earliest robust baseline schedule (indicated by return value $\KwFeasible$) or concluding that permutation $\oppermfn$ is infeasible (indicated by return value $\KwInfeasible$), i.e.\ for the given permutation $\oppermfn$, it is not possible to find a robust baseline start time for some operation.

The algorithm for computing the earliest robust baseline start time of $\opperm{\opposmax}$ is presented in the following subsections.

\subsubsection{Na{\"\i}ve algorithm for computing the earliest robust baseline start time of $\opperm{\opposmax}$}
From now on, assume that $\basesched$ is a baseline schedule where operations $\opperm{1},\opperm{2},\dots,\opperm{\opposmax-1}$ start at their earliest robust baseline start time, and we want to find the earliest robust baseline start time for $\opperm{\opposmax}$ (all other operations are not yet assigned to any start time).
A na{\"\i}ve algorithm (see Algorithm~\ref{alg:optsched/naive}) directly applies the definition of the earliest robust baseline start time: iterate over every possible baseline start time \hbox{$\basestart{\opperm{\opposmax}} \in \intinterval{\max(\releasetime{\opperm{\opposmax}}, \basestart{\opperm{\opposmax-1}} + \proctime{\opperm{\opposmax-1}})}{\basestartmax}$} in increasing order and select the earliest baseline start time such that Eq.~\eqref{eq:optsched/robust-start-time} is not violated.
However, such algorithm is inefficient since the number of realised schedules of $\basesched$ is exponential in $\numops$.

\begin{algorithm}[H]
  \caption{Earliest Robust Baseline Start Time of $\opperm{\opposmax}$ for Fixed Permutation: na{\"\i}ve version}
\label{alg:optsched/naive}
  \Fn{}
  \KwComputeEarliestRobustStartTime{$\oppermfn, \opposmax, \basesched$}
  \BeginAlgorithm{
    \DontPrintSemicolon{}
    \eIf{$\opposmax=1$}{
      $\basestart{\opperm{\opposmax}}\gets \releasetime{\opperm{\opposmax}}$\;
    }{
      $\basestart{\opperm{\opposmax}}\gets \max(\releasetime{\opperm{\opposmax}}, \basestart{\opperm{\opposmax-1}} + \proctime{\opperm{\opposmax-1}})$\;
    }
    \While{$\basestart{\opperm{\opposmax}} \le \basestartmax$}{
      $\textit{energyLimitViolated} \gets \False$\;
      \ForEach{$(\devsit, \metint) \in \devsitsset \times \metintsset$}{
\label{line:optsched/naive/begin-test}
        $\realsched \gets \realschedfn[\basesched][\devsit]$\;
        \If{$\sum_{\oppos = 1}^{\opposmax}\metintproctimesym(\metint,\opperm{\oppos},\realstart{\opperm{\oppos}}) \cdot \power{\opperm{\oppos}}>\maxenergy[\metint]$}{
          $\textit{energyLimitViolated} \gets \True$\;
          \Break{}
\label{line:optsched/naive/end-test}
        }
      }
      \eIf{$\textit{energyLimitViolated} = \False$}{
        \Break{}
      }{
        $\basestart{\opperm{\opposmax}} \gets \basestart{\opperm{\opposmax}} + 1$\;
\label{line:optsched/naive/inc}
      }
    }
    \eIf{$\basestart{\opperm{\opposmax}} > \basestartmax$}{
      \Return{\KwInfeasible}\;
    }{
      \Return{\KwFeasible}\;
    }
  }
\end{algorithm}

\subsubsection{Increasing the efficiency of the na{\"\i}ve algorithm: energy consumption dominance of the right-shift schedules}
\begin{figure}[H]
  \centering
  \includegraphics[scale=0.7]{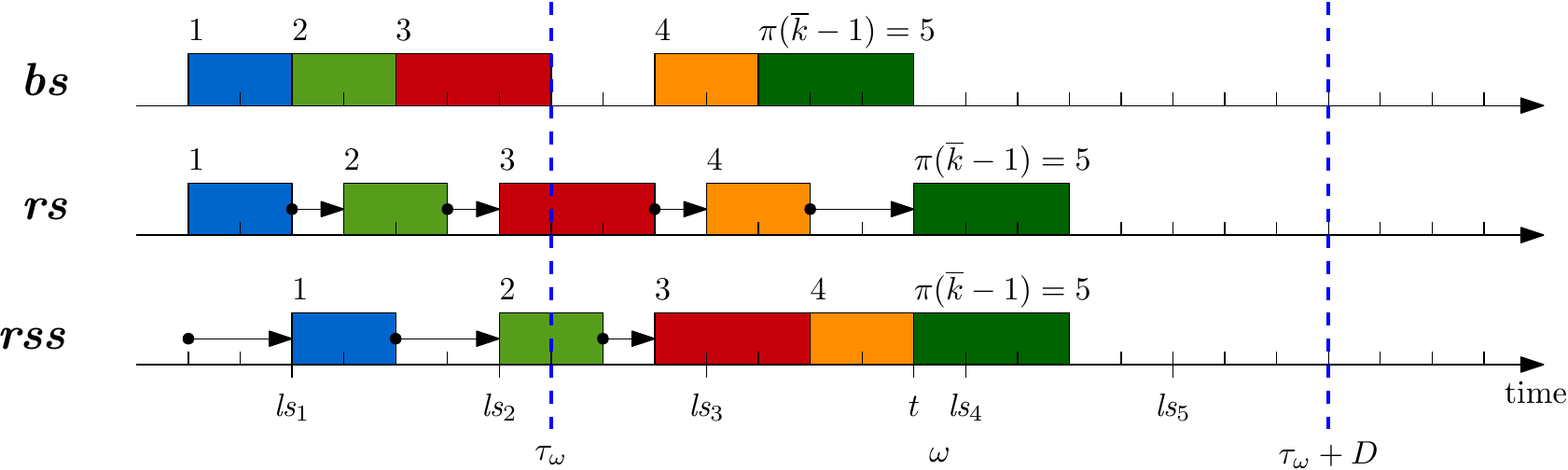}
  \caption{The right-shift schedule maximises the energy consumption, $\devmax = 2$.}
\label{fig:optsched/rss-max-energy}
\end{figure}

The first key observation for obtaining an efficient algorithm is the ``energy consumption dominance'' of the right-shift schedules, which is illustrated in Fig.~\ref{fig:optsched/rss-max-energy}.
Let $\realsched$ be some realised schedule of operations $\opperm{1},\opperm{2},\dots,\opperm{\opposmax-1}$, then it can be proven that the energy consumption in the metering intervals intersected by $\opperm{\opposmax-1}$ in \hbox{$\shiftsched = \shiftschedfnsym(\basesched, \opposmax-1, \realstart{\opperm{\opposmax-1}})$} is not less than in $\realsched$, i.e.
\begin{equation}
  \metintproctimesym(\metint,\opperm{\opposmax-1},\realstart{\opperm{\opposmax-1}}) > 0 \implies \sum_{\oppos=1}^{\opposmax-1} \metintproctimesym(\metint,\opperm{\oppos},\realstart{\opperm{\oppos}}) \cdot \power{\oppos} \le \sum_{\oppos=1}^{\opposmax-1} \metintproctimesym(\metint,\opperm{\oppos},\shiftstart{\opperm{\oppos}}) \cdot \power{\oppos},\quad \forall \metint \in \metintsset
\end{equation}
Therefore, it suffices to consider only the right-shift schedules since if some energy limit is violated in $\realsched$ then it will also be violated in $\shiftsched$ (see Theorem~\ref{lem:app/shifted-dominance} in Appendix).
Lines~\ref{line:optsched/naive/begin-test}-\ref{line:optsched/naive/end-test} in Algorithm~\ref{alg:optsched/naive} can be replaced with the following pseudo-code
\begin{algorithm}[H]
  \DontPrintSemicolon{}
  \eIf{$\opposmax=1$}{
    \ForEach{$(\dev{\opperm{\opposmax}}, \metint) \in \intinterval{0}{\devmax} \times \metintsset$}{
        $\realstart{\opperm{\opposmax}} \gets \basestart{\opperm{\opposmax}} + \dev{\opperm{\opposmax}}$\;
        \If{$\metintproctimesym(\metint,\opperm{\opposmax},\realstart{\opperm{\opposmax}}) \cdot \power{\opperm{\opposmax}}>\maxenergy[\metint]$}{
          $\textit{energyLimitViolated} \gets \True$\;
          \Break{}
        }
    }
  }{
    \ForEach{$(\mt, \dev{\opperm{\opposmax}},\metint) \in \intinterval{\basestart{\opperm{\opposmax-1}}}{\latestartfn{\basesched}{\opperm{\opposmax-1}}} \times \intinterval{0}{\devmax} \times \metintsset$}{
        $\realsched \gets \shiftschedfnsym(\basesched, \opperm{\opposmax-1}, \mt)$\;
        $\realstart{\opperm{\opposmax}} \gets \max(\basestart{\opperm{\opposmax}}, \realstart{\opperm{\opposmax-1}}+\proctime{\opperm{\opposmax-1}}) + \dev{\opperm{\opposmax}}$\;
        \If{$\sum_{\oppos = 1}^{\opposmax}\metintproctimesym(\metint,\opperm{\oppos},\realstart{\opperm{\oppos}}) \cdot \power{\opperm{\oppos}}>\maxenergy[\metint]$}{
          $\textit{energyLimitViolated} \gets \True$\;
          \Break{}
        }
    }
  }
\end{algorithm}
Although such algorithm does not have exponential complexity anymore, it is still not very efficient since it asymptotically depends on the length of the horizon.

\subsubsection{Increasing the efficiency of the na{\"\i}ve algorithm: maximum possible intersection of the operations with the metering intervals}
As was noted at the end of the previous subsection, the complexity of the na{\"\i}ve algorithm depends on the length of the horizon since whenever a realised schedule violating any energy consumption limit is found, the baseline start time of $\opperm{\opposmax}$ is increased by 1, see line~\ref{line:optsched/naive/inc} in Algorithm~\ref{alg:optsched/naive}.
The question is whether it is possible to identify a range of non-robust baseline start times of $\opperm{\opposmax}$ and, therefore, ``jump'' by more than 1 time unit on line~\ref{line:optsched/naive/inc} in Algorithm~\ref{alg:optsched/naive}.
This is possible by considering \defterm{a maximum possible intersection} of operation $\opperm{\opposmax}$ with metering intervals.

Assume that $\realsched$ is some realised schedule of operations $\opperm{1},\opperm{2},\dots,\opperm{\opposmax}$ such that the energy consumption limit is violated in some metering interval $\metint \in \metintsset$.
One of the following two cases occurs:
\begin{description}
  \item[\emph{Case 1}] $\metintproctimesym(\metint,\opperm{\opposmax-1},\realstart{\opperm{\opposmax-1}}) > 0$: since the baseline start times of $\opperm{1}, \opperm{2}, \dots, \opperm{\opposmax-1}$ are robust, $\metintproctimesym(\metint,\opperm{\opposmax},\realstart{\opperm{\opposmax}}) > 0$ must hold.
    We may ask what is the maximum possible intersection of $\opperm{\opposmax}$ in $\metint$ relative to realised start times $\realsched$ of $\opperm{1},\opperm{2},\dots,\opperm{\opposmax-1}$ without violating the energy limit
    \begin{equation}
        \textit{maxPossibleIntersection}_{\metint} = \left\lfloor {\frac{\maxenergy[\metint] - \sum_{\oppos = 1}^{\opposmax-1}\metintproctimesym(\metint,\opperm{\oppos},\realstart{\opperm{\oppos}}) \cdot \power{\opperm{\oppos}}}{\power{\opperm{\opposmax}}}} \right\rfloor
    \end{equation}
    It can be proven (see Lemma~\ref{lem:app/earlier-violates-relative} in Appendix) that all baseline start times \hbox{$\intinterval{\max(\releasetime{\opperm{\opposmax}}, \basestart{\opperm{\opposmax-1}}+\proctime{\opperm{\opposmax-1}})}{\metintend{\metint} - \textit{maxPossibleIntersection}_{\metint} - 1}$} of $\opperm{\opposmax}$ are not robust, i.e. $\metintend{\metint} - \textit{maxPossibleIntersection}_{\metint} \le \basestart{\opperm{\opposmax}}$ must hold, otherwise $\maxenergy[\metint]$ is violated.

  \item[\emph{Case 2}] $\metintproctimesym(\metint,\opperm{\opposmax-1},\realstart{\opperm{\opposmax-1}}) = 0$: in this case, $\opperm{\opposmax}$ is the only operation having a non-zero intersection with $\metint$ in $\realsched$.
  Therefore, it holds that $\metintproctimesym(\metint,\opperm{\opposmax},\realstart{\opperm{\opposmax}}) \cdot \power{\opperm{\opposmax}} > \maxenergy[\metint]$.
    We can compute the maximum possible intersection of $\opperm{\opposmax}$ with $\metint$ 
    \begin{equation}
      \textit{maxPossibleIntersection}_{\metint} = \left\lfloor\frac{\maxenergy[\metint]}{\power{\opperm{\opposmax}}}\right\rfloor
    \end{equation}
    which represents the maximum intersection length between $\metint$ and $\opperm{\opposmax}$ without violating the energy consumption limit.
    Assuming that all baseline start times $\intinterval{\max(\releasetime{\opperm{\opposmax}}, \basestart{\opperm{\opposmax-1}}+\proctime{\opperm{\opposmax-1}})}{\basestart{\opperm{\opposmax}} -1}$ of $\opperm{\opposmax}$ are not robust, it is easy to see that \hbox{$\metintend{\metint} - \textit{maxPossibleIntersection}_{\metint} \le \basestart{\opperm{\opposmax}}$} must hold to assure that the intersection of $\opperm{\opposmax}$ with $\metint$ in any realised schedule is not larger than $\textit{maxPossibleIntersection}_{\metint}$.
\end{description}

Due to these two cases, the computation of the earliest robust baseline start time can be split into two consecutive steps (corresponding to the cases described above)
\begin{description}
  \item[\emph{Step 1}] the earliest robust baseline start time relative to $\opperm{1},\opperm{2},\dots,\opperm{\opposmax-1}$: for each right-shift schedule $\shiftschedfnsym(\basesched, \opperm{\opposmax-1}, \mt)$, where $\mt \in \intinterval{\basestart{\opperm{\opposmax-1}}}{\latestartfn{\basesched}{\opperm{\opposmax-1}}}$, find the earliest baseline start time of $\opperm{\opposmax}$ using the maximum possible intersection which does not violate $\maxenergy[\metint]$, where $\metint \in \metintsset$ is the last metering interval having a non-zero intersection with $\opperm{\opposmax-1}$.
    Notice, that due to Lemma~\ref{lem:app/earlier-violates-relative} it is efficient to check $\mt$ in decreasing order since if some energy consumption limit is violated for some $\realstart{\opperm{\opposmax}}$, then all baseline start times $\basestart{\opperm{\opposmax}} \le \realstart{\opperm{\opposmax}}$ cannot be robust and the algorithm can continue directly with the second step.

  \item[\emph{Step 2}] the earliest robust baseline start time relative to only $\opperm{\opposmax}$: After the first step, it is easy to see that for every baseline start time $\intinterval{\basestart{\opperm{\opposmax}}}{\basestartmax}$ of $\opperm{\opposmax}$ there is no realised schedule of operation $\opperm{1},\opperm{2},\dots,\opperm{\opposmax}$ in which both $\opperm{\opposmax-1},\opperm{\opposmax}$ have a non-zero intersection with some metering interval $\metint \in \metintsset$ and $\maxenergy[\metint]$ would be violated.
    However, the energy limits can still be violated in metering intervals in which only $\opperm{\opposmax}$ have a non-zero intersection.

    Consider the example from Fig.~\ref{fig:optsched/itself}.
    For each $\mt \in \intinterval{\basestart{\opperm{\opposmax}}}{\latestart{\opperm{\opposmax}}}$, there exists a realised schedule $\realsched$ of operations $\opperm{1},\opperm{2},\dots,\opperm{\opposmax}$ such that $\realstart{\opperm{\opposmax}} = \mt$.
    Therefore, the maximum intersection of $\opperm{\opposmax}$ with every metering interval $\metint \ge \left\lfloor \frac{\basestart{\opperm{\opposmax}}}{\lenmetint} \right\rfloor$ is
    \begin{equation}
      \textit{maxIntersection}_{\metint} = \min(\proctime{\opperm{\opposmax}}, \text{lenint}([\metintstart{\metint},\metintend{\metint}],[\basestart{\opperm{\opposmax}},\latestart{\opperm{\opposmax}} + \proctime{\opperm{\opposmax}}]))
    \end{equation}
    Then, $\basestart{\opperm{\opposmax}}$ is not robust if there exists metering interval $\metint$ such that the maximum intersection is larger than the maximum possible intersection in $\metint$, i.e.
    \begin{equation}
      \textit{maxPossibleIntersection}_{\metint} = \left\lfloor\frac{\maxenergy[\metint]}{\power{\opperm{\opposmax}}}\right\rfloor < \textit{maxIntersection}
    \end{equation}
    If this is the case, then the earliest baseline start time that can be robust is \hbox{$\metintend{\metint} - \textit{maxPossibleIntersection}_{\metint}$}.
    \begin{figure}[H]
      \centering
      \includegraphics[scale=0.7]{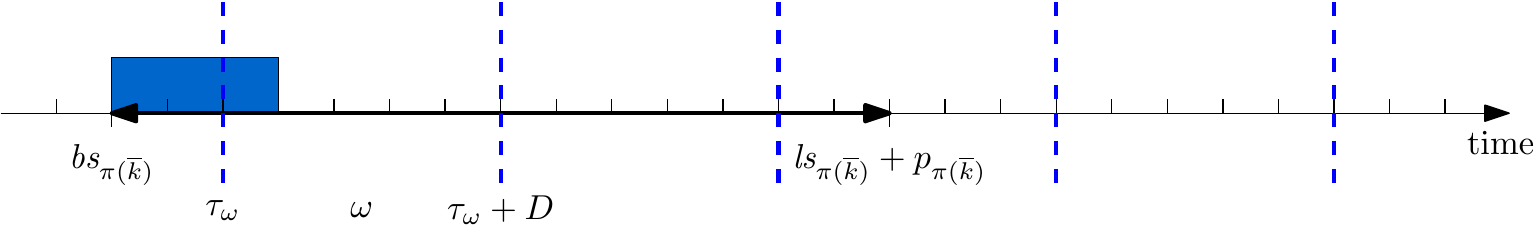}
      \caption{The time interval intersected by operation $\opperm{\opposmax}$.}
\label{fig:optsched/itself}
    \end{figure}
\end{description}

The complete algorithm for computing the robust baseline start time that combines all the discussed ideas is shown in Algorithm~\ref{alg:optsched/optimised}.

\subsection{Algorithm complexity of the algorithm for computing the robust baseline schedule}
\label{sec:optsched/complete}
The complexity of \emph{Step 1} of Algorithm~\ref{alg:optsched/optimised} is $\acupperbound{\numops^2 \cdot \devmax}$ since the number of unique right-shift schedules of $\opperm{\opposmax-1}$ is bounded above by $\numops \cdot \devmax$ and the computing a right-shift schedule can be done in $\acupperbound{\numops}$.
The complexity of \emph{Step 2} is $\acupperbound{|\metintsset|}$.
Since Algorithm~\ref{alg:optsched/optimised} is repeated for each position in the permutation (see Algorithm~\ref{alg:optsched/schedule}), the complexity of computing the optimal robust schedule for the given permutation is $\acupperbound{\numops^3 \cdot \devmax + \numops \cdot |\metintsset|}$; the pseudo-polynomiality of the algorithm arises due to term $\devmax$.
Notice that if the operations cannot violate the energy consumptions limits by themselves, i.e. \hbox{$\forall \metint \in \metintsset, \forall \op \in \opsset: \min(\proctime{\op}, \lenmetint) \cdot \power{\op} \le \maxenergy[\metint]$}, then \emph{Step 2} is not necessary and the complexity is $\acupperbound{\numops^3 \cdot \devmax}$.
\newpage

\begin{algorithm}[H]
  \caption{Earliest Robust Baseline Start Time of $\opperm{\opposmax}$ for Fixed Permutation}
\label{alg:optsched/optimised}
  \Fn{}
  \KwComputeEarliestRobustStartTime{$\oppermfn, \opposmax, \basesched, \latesched$}
  \BeginAlgorithm{
    \DontPrintSemicolon{}
    $\basestart{\opperm{\opposmax}} \gets \releasetime{\opperm{\opposmax}}$\;
    \tcc{Step 1: Earliest robust baseline start time relative to $\opperm{1},\opperm{2},\dots,\opperm{\opposmax-1}$.}
    \If{$\opposmax > 1$} {
      $\basestart{\opperm{\opposmax}} \gets \max(\releasetime{\opperm{\opposmax}}, \basestart{\opperm{\opposmax-1}} + \proctime{\opperm{\opposmax-1}})$\;
      $\mt \gets \latestart{\opperm{\opposmax - 1}}$\;
      \While{$\mt \ge \min(\latestart{\opperm{\opposmax - 1}},\max(\basestart{\opperm{\opposmax - 1}}, \releasetime{\opperm{\opposmax}}-\proctime{\opperm{\opposmax-1}}))$}{
\label{algline:optsched/basic/eas-due-preceding-start}
        $\shiftsched \gets \shiftschedfnsym(\basesched, \opperm{\opposmax-1}, \mt)$\;
        $\metint \gets \left\lfloor\frac{\shiftstart{\opperm{\opposmax - 1}} + \proctime{\opperm{\opposmax - 1}} - 1}{\lenmetint}\right\rfloor$\;
        $\textit{maxPossibleIntersection} = \left\lfloor {\frac{\maxenergy[\metint] - \sum_{\oppos = 1}^{\opposmax-1}\metintproctimesym(\metint,\opperm{\oppos},\shiftstart{\opperm{\oppos}}) \cdot \power{\opperm{\oppos}}}{\power{\opperm{\opposmax}}}} \right\rfloor$\;
        \uIf{$\proctime{\opperm{\opposmax}} \le \textit{maxPossibleIntersection}$}{
          \tcc{Small optimisation: Entire operation can be fitted into $\metint$ without violating $\maxenergy[\metint]$, therefore continue with metering interval $\metint -1$.}
          $\mt \gets \metintstart{\metint} - \proctime{\opperm{\opposmax-1}} - 1$\;
        }
        \uElseIf{$\textit{maxPossibleIntersection} \ge (\metintend{\metint}) - (\shiftstart{\opperm{\opposmax - 1}} + \proctime{\opperm{\opposmax - 1}})$}{
          \tcc{$\maxenergy[\metint]$ is not violated.}
            $\mt \gets \mt - 1$\;
        }
        \uElse{
          \tcc{$\maxenergy[\metint]$ is violated.}
          $\basestart{\opperm{\opposmax}} \gets \max(\releasetime{\opperm{\opposmax}}, (\metintend{\metint}) - \textit{maxPossibleIntersection})$\;
          \Break{}
        }
      }
    }
    \tcc{Step 2: Earliest robust baseline start time relative to only $\opperm{\opposmax}$.}
    $\latestart{\opperm{\opposmax}} \gets \latestartfn{\basesched}{\opperm{\opposmax}}$\;
    $\metint \gets \left\lfloor \frac{\basestart{\opperm{\opposmax}}}{\lenmetint} \right\rfloor$\;
    \While{$\metint \in \metintsset$}{
      $\textit{maxPossibleIntersection} \gets \left\lfloor\frac{\maxenergy[\metint]}{\power{\opperm{\opposmax}}}\right\rfloor$\;
      $\textit{maxIntersection} \gets \min(\proctime{\opperm{\opposmax}}, \text{lenint}([\metintstart{\metint},\metintend{\metint}],[\basestart{\opperm{\opposmax}},\latestart{\opperm{\opposmax}} + \proctime{\opperm{\opposmax}}]))$\;
      \uIf{$\textit{maxIntersection} = 0$}{
        \Break{}
      }
      \ElseIf{$\textit{maxPossibleIntersection} < \textit{maxIntersection}$}{
        $\basestart{\opperm{\opposmax}} \gets \metintend{\metint} - \textit{maxPossibleIntersection}$\;
        $\latestart{\opperm{\opposmax}} \gets \latestartfn{\basesched}{\opperm{\opposmax}}$\;
      }
      $\metint \gets \metint + 1$\;
    }
    \eIf{$\basestart{\opperm{\opposmax}} > \basestartmax$}{
      \Return{\KwInfeasible}\;
    }{
      \Return{\KwFeasible}\;
    }
  }
\end{algorithm}

\newpage


\section{Exact Algorithms for $\grahamwhole$}
\label{sec:exact}
We propose two exact algorithms for $\grahamwhole$: (i) Branch-and-Bound and (ii) a logic-based Benders decomposition algorithm.
Both algorithms exploit the fact that the optimal robust schedule for the fixed permutation can be found by Algorithm~\ref{alg:optsched/schedule}.

\subsection{Branch-and-Bound}
\label{sec:exact/bb}
Since robustness of a schedule with fixed operations' order can be solved independently by Algorithm~\ref{alg:optsched/schedule}, the proposed Branch-and-Bound algorithm (BB) is very simple.
The algorithm searches through the space of partial permutations, i.e.\ in each node of the search tree, BB constructs the earliest robust schedule for partial permutation $\opperm{1},\opperm{2}, \dots, \opperm{\opposmax-1}$ using Algorithm~\ref{alg:optsched/schedule}.
For each remaining operation \hbox{$\opperm{\opposmax} \in \opsset \setminus \{\opperm{1},\opperm{2}, \dots, \opperm{\opposmax-1}\}$}, a new branch is created by appending operation $\opperm{\opposmax}$ to the end of the current partial permutation $\oppermfn$, i.e.\ the set of new branches is
\begin{equation}
  \{\opperm{1},\opperm{2}, \dots, \opperm{\opposmax-1},\opperm{\opposmax} \mid \opperm{\opposmax} \in \opsset \setminus \{\opperm{1},\opperm{2}, \dots, \opperm{\opposmax-1}\} \}
\end{equation}
Our BB is implemented as Depth-First Search, and the branches are prioritised according to the due dates of $\opsset \setminus \{\opperm{1},\opperm{2}, \dots, \opperm{\opposmax-1}\}$.

The solution space is pruned using Chu's lower bound~\cite{chu1992}.
The lower bound is designed for problem $1|\releasetime{\op}|\sum \tardiness{\op}$ and can be computed as follows.
W.l.o.g.\ assume that the due dates of operations are sorted, i.e. $\duedate{1} \le \duedate{2} \le \dots \le \duedate{\numops}$.
Schedule the operations according to the Shortest Remaining Processing Time rule (i.e.\ the operations can be preempted); let $\completionord{\oppos}$ be the completion time in the baseline schedule of $\oppos$-th completed operation.
Then Chu's lower bound on the total tardiness is $\sum_{\oppos=1}^{\numops} \max(\completionord{\oppos} - \duedate{\oppos}, 0)$.

\subsection{Logic-based Benders decomposition algorithm}
\label{sec:exact/lbbd}
\defterm{Logic-based Benders decomposition} (LBBD)~\cite{hooker2007} is a generalisation of the classical Benders decomposition that is used for solving large-scale optimisation problems.
In the classical Benders decomposition, the subproblem is only a continuous linear or non-linear problem whereas in LBBD the subproblem may have an arbitrary form.
We use a specific form of the LBBD in which the cuts remove infeasible solutions (\defterm{no-good} cuts).

The idea of LBBD is to decompose the original problem into two parts: (i) \defterm{master problem}, which is a relaxation of the original problem and (ii) \defterm{subproblem}.
After the master problem is solved to optimality, its solution is checked by the subproblem whether it is feasible in the original problem or not.
If yes, then the decomposition algorithm finishes since an optimal feasible solution for the original problem has been found.
If not, a \defterm{cut} constraint is generated in which the infeasible solution is violated.
The cut is added to the master problem, and the whole procedure is repeated.

In modern implementations of LBBD, the cuts are added gradually during solving the master problem~\cite{lin2016}.
This approach is more integrated into an Mixed Integer Linear Programming (MILP) solvers and therefore more efficient since the master problem does not need to be resolved from scratch every time a new cut is generated; the state-of-the-art solvers such as Gurobi or CPLEX support adding cuts dynamically using \defterm{lazy constraints generation} mechanism.

In our case, the master problem is essentially a MILP model of $\grahamenergylimits$.
The solution of the master problem, i.e.\ baseline schedule $\baseschedmm$, is checked in the subproblem whether it is robust or not.
If $\baseschedmm$ is not robust, a no-good cut is generated for the master problem.

\subsubsection{Master Problem}
\label{sec:exact/master}
The MILP formulation of the master problem corresponds to Constraints~\eqref{eq:exact/master/starts-once}-\eqref{eq:exact/master/single-dev} and objective~\eqref{eq:exact/master/objective}.
It is modelled as a time-indexed formulation, which is suitable for generating the cuts.
There are two types of variables in the program: (i) a \defterm{binary baseline start time} of operation $\op$ in time $\mt$ defined as $\basestartbin{\op}{\mt} = 1$ iff $\op$ starts at $\mt$ in the baseline schedule and (ii) the \defterm{energy consumed in time} $\mt$ denoted as $\timeenergy{\mt}$.
\begin{align}
  \label{eq:exact/master/objective}
  &\min \enspace \sum_{\op \in \opsset} \sum_{\mt = \releasetime{\op}}^{\basestartmax} \basestartbin{\op}{\mt} \cdot \max(0, \mt + \proctime{\op} - \duedate{\op})  \\
  \label{eq:exact/master/starts-once}
  \mbox{s.t. }&\sum_{\mt = \releasetime{\op}}^{\basestartmax} \basestartbin{\op}{\mt} =  1 \enspace,\quad\quad  \op \in \opsset \\
  \label{eq:exact/master/alloc-one}
  &\sum_{\op \in \opsset} \sum_{\mt' = \max(\releasetime{\op}, \mt - \proctime{\op} + 1)}^{\min(\mt, \basestartmax)} \basestartbin{\op}{\mt'} \le 1 \enspace,\quad\quad  \mt \in \intinterval{\minreleasetime}{\horizon - 1} \\
  \label{eq:exact/master/time-energy}
  &\sum_{\op \in \opsset} \sum_{\mt' = \max(\releasetime{\op}, \mt - \proctime{\op} + 1)}^{\min(\mt, \basestartmax)} \basestartbin{\op}{\mt'} \cdot \power{\op}= \timeenergy{\mt} \enspace,\quad\quad  \mt \in \intinterval{\minreleasetime}{\horizon - 1} \\
  \label{eq:exact/master/non-deviated-limits}
  &\sum_{\mt = \metintstart{\metint}}^{\metintend{\metint} - 1 } \timeenergy{\mt} \le \maxenergy[\metint] \enspace,\quad\quad \metint \in \metintsset \\
  \label{eq:exact/master/single-dev}
  &\sum_{\mt = \max(0, \metintstart{\metint} - \devsym)}^{\metintend{\metint} - 1 - \devsym} \timeenergy{\mt} \le \maxenergy[\metint] ,\enspace\quad \forall \devsym \in \intinterval{1}{\devmax}, \forall \metint \in \metintsset
\end{align}

The objective~\eqref{eq:exact/master/objective} of the master problem is the minimisation of $\sum_{\op \in \opsset} \tardiness{\op}$.
Constraint~\eqref{eq:exact/master/starts-once} ensures that each operation starts in some time that is at least its release time and at most the maximum start time $\basestartmax$.
Constraint~\eqref{eq:exact/master/alloc-one} enforces that each time can be occupied by at most one operation.
Computation of consumed energy in time $\mt$ is in Constraint~\eqref{eq:exact/master/time-energy}.
\hbox{Constraint~\eqref{eq:exact/master/non-deviated-limits}} ensures that the energy consumption limit in each metering interval is not violated for the baseline schedule.
Finally, Constraint~\eqref{eq:exact/master/single-dev} strengthens the master problem by taking into account a subset of scenarios in which only a single operation deviated.
Although this constraint is valid for problem $\grahamwhole$, it is not exact, i.e.\ not every infeasible realised schedule, in which one operation deviated, is cut out.
For more details about Constraint~\eqref{eq:exact/master/single-dev}, see~\cite{modos2016a}.

\subsubsection{Subproblem: robustness check and cuts}
When integer baseline schedule $\baseschedmm$ is found by the MILP solver, it is checked if it is robust, i.e.\ whether all Constraints~\eqref{eq:prob/robust/robust} are satisfied.
If the schedule is not robust, a no-good cut is generated for the master problem such that $\baseschedmm$ violates it.
For checking the robustness of $\baseschedmm$, a pseudo-polynomial algorithm introduced in~\cite{modos2016a} is employed.


Now assume that schedule $\baseschedmm$ is not robust.
In general, our cuts have a form of
\begin{equation}
  \label{eq:exact/subproblem/cutting-constraint}
  \sum_{\op \in \opsset} \sum_{\mt \in \cutintmm{\op}} \basestartbin{\op}{\mt} \le \numops - 1
\end{equation}
where $\cutintmm{\op}$ is a \defterm{cutting interval} of operation $\op$ for schedule $\baseschedmm$, i.e.\ the cut enforces that at least one operation starts outside of its cutting interval.

Simple cutting intervals that forbid one particular schedule $\baseschedmm$ can be defined as $\forall \op \in \opsset: \cutintmm{\op} = \{\basestartmm{\op}\}$.
In the current work, we introduce cutting intervals that exploit the knowledge of the optimal robust schedule $\baseschedoptmm$ obtained by Algorithm~\ref{alg:optsched/schedule} for permutation $\oppermfnmm$ corresponding to schedule $\baseschedmm$.
Informally, the cutting intervals are defined in such a way that the start times of the operations in the baseline schedules having the same order as in $\oppermfnmm$ are ``pushed'' towards $\baseschedoptmm$.

The type of the generated cuts depends on the return value of Algorithm~\ref{alg:optsched/schedule} for permutation $\oppermfnmm$
\begin{enumerate}
\item $\KwInfeasible, \emptyset$: this means that any baseline schedule having the same order as $\oppermfnmm$ cannot be robust.
  Therefore, a cut must be generated that ``forbids'' $\oppermfnmm$.
    Such cut can be formulated using binary precedence variables $y_{\op, \op'}$~\cite{kone2011} such that $y_{\op, \opprime} = 1$ if operation $\op$ precedes operation $\opprime$, 0 otherwise.
    However, to model these variables, Big M constraints are usually employed which have poor relaxation.
    Alternatively, we can forbid only a subset of all schedules having the same order as $\oppermfnmm$ with a simpler constraint described below.

    Consider two operations $\oppermmm{\oppos}, \oppermmm{\oppos + 1}$ that are executed consecutively and let $\basestartmm{\oppermmm{\oppos}}, \basestartmm{\oppermmm{\oppos + 1}}$ be their baseline start times, respectively.
    Consider another schedule $\basesched[\prime\prime]$ in which the start time of $\oppermmm{\oppos + 1}$ is at least $\basestartmm{\oppermmm{\oppos + 1}}$.
    Then to guarantee that $\oppermmm{\oppos}$ starts before $\oppermmm{\oppos + 1}$ in $\basesched[\prime\prime]$, the start time of $\oppermmm{\oppos}$ must be at most $\basestartmm{\oppermmm{\oppos + 1}} + \proctime{\oppermmm{\oppos + 1}} - 1$.
    Therefore, in any integer schedule $\basesched$ such that
    \begin{align}
      \basestart{\oppermmm{\oppos}} &\in \intinterval{\basestartmm{\oppermmm{\oppos}}}{\min(\basestartmm{\oppermmm{\oppos + 1}} + \proctime{\oppermmm{\oppos + 1}} - 1, \basestartmax)} \\
      \basestart{\oppermmm{\oppos + 1}} &\in \intinterval{\basestartmm{\oppermmm{\oppos + 1}}}{\basestartmax}
    \end{align}
    operation $\oppermmm{\oppos}$ is executed before $\oppermmm{\oppos + 1}$.
    
    Such intervals can be derived for whole permutation $\oppermfnmm$.
    Therefore, the cutting intervals are
    \begin{equation}
      \cutintmm{\oppermmm{\oppos}} =
      \begin{dcases}
        \intinterval{\basestartmm{\oppermmm{\oppos}}}{\min(\basestartmm{\oppermmm{\oppos + 1}} + \proctime{\oppermmm{\oppos + 1}} - 1, \basestartmax)} & \oppos \in \intinterval{1}{\numops - 1} \\
        \intinterval{\basestartmm{\oppermmm{\oppos}}}{\basestartmax} & \oppos = \numops
      \end{dcases}
    \end{equation}
    It is guaranteed that if all the operations start anywhere in these intervals, the order of operations is the same as in infeasible permutation $\oppermfnmm$.
    
  \item $\KwFeasible,\baseschedoptmm$: since $\baseschedmm \not= \baseschedoptmm$ (otherwise $\baseschedmm$ would be robust), there exists position $\opposmax$ in permutation $\oppermfnmm$ such that
    \begin{align}
    &\basestartmm{\oppermmm{\oppos}} = \basestartopt{\oppermmm{\oppos}}\,,\quad \oppos \in \intinterval{1}{\opposmax - 1} \\
    &\basestartmm{\oppermmm{\opposmax}} \not= \basestartopt{\oppermmm{\opposmax}}
    \end{align}

    One of the following two cases occurs
    \begin{enumerate}
        \item $\basestartmm{\oppermmm{\opposmax}} < \basestartopt{\oppermmm{\opposmax}}$:
          Consider the example in Fig.~\ref{fig:exact/exact-subproblem-less} illustrating this case
          \begin{figure}[H]
            \centering
            \includegraphics[scale=0.7]{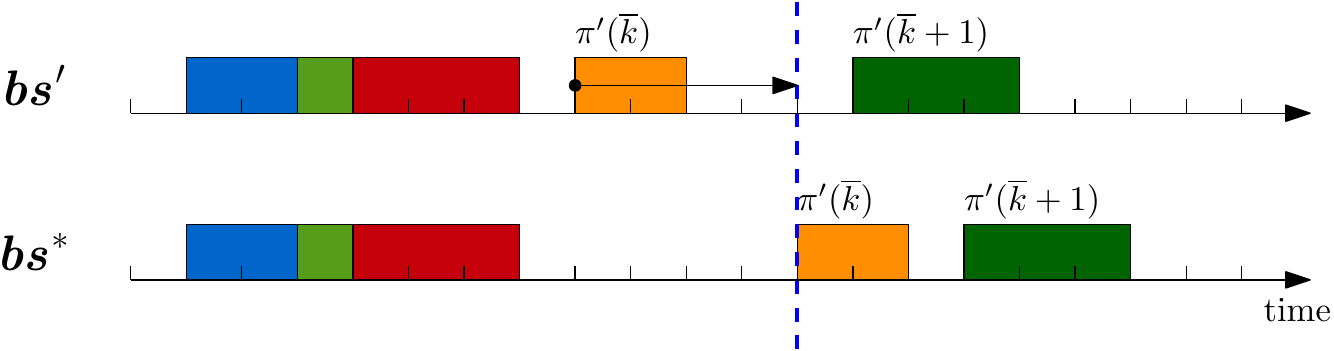}
            \caption{Case $\basestartmm{\oppermmm{\opposmax}} < \basestartopt{\oppermmm{\opposmax}}$.}
\label{fig:exact/exact-subproblem-less}
          \end{figure}

          The idea of the cut is that if in any baseline schedule $\basesched$ the operations on positions $1, \dots, \opposmax$ are the same as in $\oppermfnmm$, i.e.
          \begin{equation}
            \opperm{\oppos} = \oppermmm{\oppos}, \quad \forall \oppos \in \intinterval{1}{\opposmax},
          \end{equation}
          then the start time of $\oppermmm{\opposmax}$ should be pushed towards $\basestartopt{\oppermmm{\opposmax}}$ since all baseline start times before $\basestartopt{\oppermmm{\opposmax}}$ are not robust for $\oppermmm{\opposmax}$.
          Since the earliest robust baseline start time of $\oppermmm{\opposmax}$ is by definition dependent only on positions $\oppos \le \opposmax$ (see Eq.~\eqref{eq:optsched/robust-start-time}), the order of the operations on positions $\oppos > \opposmax$ is not important, the cut only has to guarantee that they are not executed before $\oppermmm{\opposmax}$.
          Moreover, we have to take care of the case when some operation $\{\oppermmm{\oppos} \mid \oppos \in \intinterval{\opposmax + 1}{\numops}\}$ starts before $\basestartopt{\oppermmm{\opposmax}}$ in $\baseschedmm$; to make sure that feasible schedule $\basesched$ in which some operation $\{\oppermmm{\oppos} \mid \oppos \in \intinterval{\opposmax + 1}{\numops}\}$ starts before $\oppermmm{\opposmax}$ is not cut out, $\basestartmm{\oppermmm{\opposmax + 1}}$ bounds from above the cutting interval of $\opperm{\opposmax}$ and from below the cutting intervals of $\{\oppermmm{\oppos} \mid \oppos \in \intinterval{\opposmax + 1}{\numops}\}$.
          Therefore, the cutting intervals are
          \begin{equation}
            \cutintmm{\oppermmm{\oppos}} =
            \begin{dcases}
              \intinterval{\basestartmm{\oppermmm{\oppos}}}{\min(\basestartmm{\oppermmm{\oppos + 1}} + \proctime{\oppermmm{\oppos + 1}} - 1, \basestartmax)} & \oppos \in \intinterval{1}{\opposmax - 1} \\
              \intinterval{\basestartmm{\oppermmm{\oppos}}}{\basestartopt{\oppermmm{\oppos}} - 1} & \oppos = \opposmax \,\wedge \,\opposmax = \numops \\
              \intinterval{\basestartmm{\oppermmm{\oppos}}}{\min(\basestartopt{\oppermmm{\oppos}}, \basestartmm{\oppermmm{\opposmax + 1}} + \proctime{\oppermmm{\opposmax + 1}}) - 1} & \oppos = \opposmax \,\wedge \,\opposmax < \numops \\
              \intinterval{\min(\basestartopt{\oppermmm{\opposmax}}, \basestartmm{\oppermmm{\opposmax + 1}} + \proctime{\oppermmm{\opposmax + 1}} - \proctime{\oppermmm{\oppos}})}{\basestartmax} & \oppos \in \intinterval{\opposmax + 1}{\numops}
            \end{dcases}
          \end{equation}

        \item $\basestartmm{\oppermmm{\opposmax}} > \basestartopt{\oppermmm{\opposmax}}$: this case is analogous to the previous one with the difference that operation $\oppermmm{\opposmax}$ is being pushed to the left to $\basestartopt{\oppermmm{\opposmax}}$
          \begin{equation}
            \cutintmm{\oppermmm{\oppos}} =
            \begin{dcases}
              \intinterval{\basestartmm{\oppermmm{\oppos}}}{\min(\basestartmm{\oppermmm{\oppos + 1}} + \proctime{\oppermmm{\oppos + 1}} - 1, \basestartmax)} & \oppos \in \intinterval{1}{\opposmax - 1} \\
              \intinterval{\basestartopt{\oppermmm{\oppos}} + 1}{\basestartmm{\oppermmm{\oppos}}} & \oppos = \opposmax \\
              \intinterval{\basestartmm{\oppermmm{\opposmax}} - \proctime{\oppermmm{\oppos}} + 1}{\basestartmax} & \oppos \in \intinterval{\opposmax + 1}{\numops}
            \end{dcases}
          \end{equation}
    \end{enumerate}
\end{enumerate}

Notice, that even though the cutting intervals may cut out robust baseline schedules in which the operations are not starting at their earliest robust start times, they do not cut out the optimal schedule $\baseschedoptmm$.
Therefore, the lazy constraints approach is exact.

\section{Heuristic Algorithm for $\grahamwhole$}
\label{sec:heur}
Since exact approaches are not able to provide solutions in a reasonable time for larger instances, the optimality is often sacrificed for efficiency in practical applications.
In this section, we provide a heuristic that exploits the fact that for a given permutation, the optimal robust schedule can be found by Algorithm~\ref{alg:optsched/schedule}.
Therefore, it is enough to search through the space of distinct permutations instead of a larger space of the baseline start times.

The heuristic has two-stages: construction of the initial permutation using a greedy algorithm and a tabu search algorithm that improves the initial permutation.

\subsection{Greedy algorithm for initial permutation}
\label{sec:heur/greedy}
The initial permutation is constructed in a greedy way using Algorithm~\ref{alg:heur/initperm}.
In each iteration $\opposmax$ of the algorithm, some operation from the set of not assigned operations $\textit{notAssignedOperations}$ is assigned to position $\opposmax$ in the permutation.
The algorithm selects such operation $\op \in \textit{notAssignedOperations}$ which minimises the lower bound on the total tardiness of the operations in $\textit{notAssignedOperations}$.
To compute the lower bound, the baseline start time of operation $\op$ relative to the current partial permutation is found by Algorithm~\ref{alg:optsched/optimised} and the remaining operations \hbox{$\opprime \in \textit{notAssignedOperations} \setminus \{ \op \}$} are allocated to the maximum of their release time and the completion time of operation $\op$ (notice that operations \hbox{$\textit{notAssignedOperations} \setminus \{ \op \}$} may overlap).
If any two operations achieve the same value of the lower bound, the algorithm selects the operation which completes the earliest.
\begin{algorithm}[H]
  \caption{Initial permutation.}
\label{alg:heur/initperm}
  \Fn{}
  \KwInitialPermutation{}
  \BeginAlgorithm{
    \DontPrintSemicolon{}
    $\basesched \gets (\infty, \infty, \dots, \infty)$\;
    $\latesched \gets (\infty, \infty, \dots, \infty)$\;
    $\oppermfn \gets \emptyset$\;
    $\textit{notAssignedOperations} \gets \opsset$\;
    \ForEach{$\opposmax = 1, \dots,\numops$}{
      $\objval^* \gets \infty$\;
      $\textit{bestCompletionTime} \gets \infty$\;
      $\op^* \gets 0$\;
      \ForEach{$\op \in \textit{notAssignedOperations}$}{
        $\opperm{\opposmax} \gets \op$\;
        \If{$\KwComputeEarliestRobustStartTime(\oppermfn, \opposmax, \basesched, \latesched) \not= \KwInfeasible$} {
          $\textit{completionTime} \gets \basestart{\op} + \proctime{\op}$\;
          $\objval \gets \max(\textit{completionTime} - \duedate{\op}, 0)$\;
          \ForEach{$\opprime = \textit{notAssignedOperations} \setminus \{ \op \}$}{
            $\objval \gets \objval + \max(\max(\textit{completionTime}, \releasetime{\opprime}) + \proctime{\opprime} - \duedate{\opprime}, 0)$\;
          }
          \If{$\objval < \objval^* \, \vee \, \left(\objval = \objval^* \, \wedge \, \textit{completionTime} \le \textit{bestCompletionTime}\right)$} {
            $\objval^* \gets \objval$\;
            $\op^* \gets \op$\;
            $\textit{bestCompletionTime} \gets \textit{completionTime}$\;
          }
        }
      }
      \If{$\objval^* = \infty$} {
        \tcc{The partial permutation is infeasible.}
        \Return{$\emptyset$}\;
      }
      $\opperm{\opposmax} \gets \op^*$\;
      $\KwComputeEarliestRobustStartTime(\oppermfn, \opposmax, \basesched, \latesched)$\;
      $\textit{notAssignedOperations} \gets \textit{notAssignedOperations} \setminus \{\op^*\}$\;
    }
    \Return{$\oppermfn$}\;
  }
\end{algorithm}

\subsection{Tabu Search}
\label{sec:heur/tabu}
To improve the initial solution found by Algorithm~\ref{alg:heur/initperm}, we employ a simple tabu search~\cite{glover1989-tabu1}.
The tabu search explores the space of the permutations of the operations and the tabu list contains the previously visited permutations.
The neighbourhood of some permutation is generated randomly using two moves: swapping of two randomly selected operations and moving one randomly selected operation to another position.
The best permutation in the neighbourhood, which is not tabu, is selected as a basis for the next iteration.
The tabu search finishes after executing the predefined number of iterations. 

The parameters for the tabu search were set according to the preliminary experiments as follows
\begin{table}[H]
\begin{tabular}{cc}
  \toprule
  Parameter & Value \\
  \midrule
  Number of restarts & 5 (including the initial run) \\
  Number of iterations & 200 \\
  Neighbourhood size & 50 \\
  Tabu list length & 5 \\
  \bottomrule
\end{tabular}
\centering
\caption{Tabu search parameters.}
\label{tab:heur/tabu/params}
\end{table}


\section{Experiments}
\label{sec:exp}

The performance and efficiency of the proposed exact and heuristic algorithms were evaluated using the following two experiments: (i) comparison of the exact and heuristic algorithms on small instances relative to the obtained objective values and (ii) tabu search algorithm evaluation on large instances.
All experiments were executed on an Intel(R) Core(TM) i5--4460  CPU @ 3.20GHz computer with 8GB of RAM running Fedora 23 operating system.
Gurobi Optimizer 7.0 was used for solving the master problem (see Section~\ref{sec:exact/master}) while the rest of the algorithms were programmed in C++ and compiled with GCC 5.3.1.

The source code of the algorithms and the generated instances are publicly available at \url{https://github.com/CTU-IIG/RSECLP}.

\subsection{Fist experiment: comparison of the exact and heuristic algorithms on small instances}
\label{sec:exp/comparison}

\subsubsection{Instances}
Three different sets of instances were generated randomly for various values of $\numops \in \{5, 10, 15\}$.
The number of metering intervals and the length of the metering intervals was fixed to $3 \cdot \numops$ and $15$, respectively.
For simplicity reason, the maximum energy consumption in each metering interval is the same and was fixed to $\maxenergy = 100$.
In each instance set, three values $\alpha_1, \alpha_2, \alpha_3$ were used to control how the parameters of the instances were generated: $\proctime{\op}$ was sampled from discrete uniform distribution $\unifdistd{1}{\lenmetint}$; exponential distribution with mean $\alpha_1 \cdot \frac{\sum_{\op \in \opsset} \proctime{\op}}{\numops}$ was used to sample the interarrival time of the operations, i.e.\ the difference between the release times of two consecutive operations; for generating $\duedate{\op}$, value of $\duedate{\op} - (\releasetime{\op} + \proctime{\op})$ was sampled from $\unifdistd{0}{\left\lceil \alpha_2 \cdot \sum_{\op \in \opsset} \proctime{\op}\right\rceil}$; $\power{\op}$ was sampled from continuous uniform distribution $\unifdistc{\frac{\alpha_3 \cdot \maxenergy}{\proctime{\op}}}{\frac{\maxenergy}{\proctime{\op}}}$.
Notice that to avoid generating infeasible instances, the operations itself cannot violate the energy limit.

For each triple $(\alpha_1, \alpha_2, \alpha_3) \in \{0.6, 0.9\} \times \{0.1, 0.3\} \times \{0.1, 0.3, 0.5\}$, parameters $\proctime{\op},\power{\op},\releasetime{\op},\duedate{\op}$ were randomly sampled 10-times according to the description above.
For the sampled parameters $\proctime{\op},\power{\op},\releasetime{\op},\duedate{\op}$, three instances differing only in the maximum deviation \hbox{$\devmax \in \{0, 3, 5\}$} were generated.
Therefore, each instance set $\numops \in \{5, 10, 15\}$ consisted of \hbox{$2\cdot2\cdot3\cdot10\cdot3=360$} instances.

\subsubsection{Results}
In the following text, the evaluated exact and heuristic algorithms are denoted as follows: \emph{Greedy} for the algorithm finding the initial permutation (see Section~\ref{sec:heur/greedy}), \emph{Tabu} for the tabu search (see Section~\ref{sec:heur/tabu}), \emph{BB} for the Branch-and-Bound (see Section~\ref{sec:exact/bb}) and \emph{LBBD} for the logic-based Benders decomposition (see Section~\ref{sec:exact/lbbd}).
The time-limit given to each algorithm for solving each instance is 20 minutes.
The schedules found by \emph{Tabu} were used as an upper bound for the exact algorithms (the execution time of \emph{Tabu} is not reflected in the time-limit of the exact solvers since its running time for such small instances is negligible).

The average objective values with a standard deviation obtained by the algorithms for $\numops=5, \numops=10$ and $\numops=15$ are shown in Tables~\ref{tb:exp/agg-obj-n=5},~\ref{tb:exp/agg-obj-n=10} and~\ref{tb:exp/agg-obj-n=15}, respectively.
To make the tables more comprehensible, the instances were grouped by $\alpha_3$ and $\devmax$.
As expected, the objective value increases with the increasing maximum deviation, since idle times and different permutations of the operations are necessary to make the schedules robust.
The same observation applies for increasing $\frac{\alpha_3 \cdot \maxenergy}{\proctime{\op}}$, i.e.\ the lower bound on the power consumption of the operations.
Instances with $\numops=5$ (see Tab.~\ref{tb:exp/agg-obj-n=5}) are small enough that both \emph{LBBD} and \emph{BB} can found optimal solutions for every instance.
All instances with $\numops=10$ (see Tab.~\ref{tb:exp/agg-obj-n=10}) can still be solved optimally with \emph{BB}.
On the other, instances with $\numops=15$ (see Tab.~\ref{tb:exp/agg-obj-n=15}) are hard for both \emph{LBBD} and \emph{BB} algorithms.
\emph{LBBD} approach can solve the majority of the instances optimally if the maximum deviation is zero, which is not true for the \emph{BB}.
In all cases, we see that \emph{Tabu} is able to find very good solutions.

The average running times of the algorithms with a standard deviation for $\numops=5, \numops=10$ and $\numops=15$ are shown in Tables~\ref{tb:exp/agg-run-n=5},~\ref{tb:exp/agg-run-n=10} and~\ref{tb:exp/agg-run-n=15}, respectively.
The results for \emph{LBBD} and \emph{BB} correspond to the number of found optimal solutions.

The conclusion of the experiment is that with increasing the maximum deviation and the lower bound on the power consumption of the operations, the instances are harder to solve.
Tab.~\ref{tb:exp/agg-obj-n=15} suggests that $\numops = 15$ is a tipping point for any exact algorithm based on the MILP formulation of the master problem (see Section~\ref{sec:exact/master}) since it is not able to solve all the instances when the maximum deviation is zero (i.e.\ no cuts are necessary to be generated).

\begin{table}[H]
\begin{tabular}{cccccccc}
\toprule
&&\multicolumn{4}{c}{Objective value, average $\pm$ std}&\multicolumn{2}{c}{Proven optimality [\%]}\\
\cmidrule(l){3-6} \cmidrule(l){7-8}
$\alpha_3$&$\devmax$&Greedy&Tabu&LBBD&BB&LBBD&BB\\
\midrule
0.1&0&32.0 $\pm$ 22.7&27.2 $\pm$ 19.0&27.2 $\pm$ 19.0&27.2 $\pm$ 19.0&100.0&100.0\\
0.1&3&36.5 $\pm$ 23.3&34.1 $\pm$ 22.6&34.1 $\pm$ 22.6&34.1 $\pm$ 22.6&100.0&100.0\\
0.1&5&45.1 $\pm$ 25.4&42.0 $\pm$ 24.4&42.0 $\pm$ 24.4&42.0 $\pm$ 24.4&100.0&100.0\\
0.3&0&37.4 $\pm$ 19.1&35.0 $\pm$ 19.0&35.0 $\pm$ 19.0&35.0 $\pm$ 19.0&100.0&100.0\\
0.3&3&46.8 $\pm$ 21.7&44.5 $\pm$ 19.8&44.5 $\pm$ 19.8&44.5 $\pm$ 19.8&100.0&100.0\\
0.3&5&55.9 $\pm$ 24.8&53.6 $\pm$ 23.0&53.6 $\pm$ 23.0&53.6 $\pm$ 23.0&100.0&100.0\\
0.5&0&55.8 $\pm$ 33.4&51.2 $\pm$ 29.8&51.2 $\pm$ 29.8&51.2 $\pm$ 29.8&100.0&100.0\\
0.5&3&70.5 $\pm$ 34.4&67.5 $\pm$ 32.6&67.5 $\pm$ 32.6&67.5 $\pm$ 32.6&100.0&100.0\\
0.5&5&78.8 $\pm$ 33.0&77.0 $\pm$ 32.4&77.0 $\pm$ 32.4&77.0 $\pm$ 32.4&100.0&100.0\\
\bottomrule
\end{tabular}
\centering
\caption{Aggregated objective value, $\numops=5$}
\label{tb:exp/agg-obj-n=5}
\end{table}

\begin{table}[H]
\begin{tabular}{cccccccc}
\toprule
&&\multicolumn{4}{c}{Objective value, average $\pm$ std}&\multicolumn{2}{c}{Proven optimality [\%]}\\
\cmidrule(l){3-6} \cmidrule(l){7-8}
$\alpha_3$&$\devmax$&Greedy&Tabu&LBBD&BB&LBBD&BB\\
\midrule
0.1&0&119.8 $\pm$ 69.9&101.4 $\pm$ 59.7&101.4 $\pm$ 59.7&101.4 $\pm$ 59.7&100.0&100.0\\
0.1&3&164.5 $\pm$ 82.7&142.7 $\pm$ 69.9&142.7 $\pm$ 69.9&142.7 $\pm$ 69.9&65.0&100.0\\
0.1&5&202.6 $\pm$ 89.3&173.8 $\pm$ 79.2&173.7 $\pm$ 79.1&173.7 $\pm$ 79.1&45.0&100.0\\
0.3&0&178.9 $\pm$ 91.9&153.2 $\pm$ 78.1&153.2 $\pm$ 78.1&153.2 $\pm$ 78.1&100.0&100.0\\
0.3&3&236.8 $\pm$ 95.7&212.4 $\pm$ 89.0&212.4 $\pm$ 89.0&212.3 $\pm$ 89.0&42.5&100.0\\
0.3&5&275.4 $\pm$ 96.7&248.8 $\pm$ 94.8&248.8 $\pm$ 94.8&248.8 $\pm$ 94.8&17.5&100.0\\
0.5&0&231.1 $\pm$ 106.9&190.4 $\pm$ 90.0&190.2 $\pm$ 89.6&190.2 $\pm$ 89.6&100.0&100.0\\
0.5&3&295.8 $\pm$ 103.8&267.7 $\pm$ 103.2&267.7 $\pm$ 103.2&267.7 $\pm$ 103.2&35.0&100.0\\
0.5&5&336.6 $\pm$ 105.3&310.6 $\pm$ 109.1&310.6 $\pm$ 109.1&310.6 $\pm$ 109.1&10.0&100.0\\
\bottomrule
\end{tabular}
\centering
\caption{Aggregated objective value, $\numops=10$}
\label{tb:exp/agg-obj-n=10}
\end{table}

\begin{table}[H]
\begin{tabular}{cccccccc}
\toprule
&&\multicolumn{4}{c}{Objective value, average $\pm$ std}&\multicolumn{2}{c}{Proven optimality [\%]}\\
\cmidrule(l){3-6} \cmidrule(l){7-8}
$\alpha_3$&$\devmax$&Greedy&Tabu&LBBD&BB&LBBD&BB\\
\midrule
0.1&0&284.1 $\pm$ 163.3&223.3 $\pm$ 141.8&221.9 $\pm$ 140.7&222.7 $\pm$ 141.3&97.5&32.5\\
0.1&3&401.5 $\pm$ 170.2&332.4 $\pm$ 160.1&332.4 $\pm$ 160.1&330.8 $\pm$ 160.7&12.5&17.5\\
0.1&5&490.4 $\pm$ 197.0&392.6 $\pm$ 182.6&392.6 $\pm$ 182.6&392.1 $\pm$ 182.2&10.0&12.5\\
0.3&0&437.4 $\pm$ 176.9&337.9 $\pm$ 146.4&337.6 $\pm$ 146.0&337.9 $\pm$ 146.4&82.5&10.0\\
0.3&3&550.6 $\pm$ 176.1&493.0 $\pm$ 169.1&493.0 $\pm$ 169.1&492.8 $\pm$ 169.0&2.5&5.0\\
0.3&5&652.2 $\pm$ 189.4&583.1 $\pm$ 185.6&583.1 $\pm$ 185.6&582.8 $\pm$ 185.6&2.5&2.5\\
0.5&0&541.0 $\pm$ 214.7&445.8 $\pm$ 181.7&445.1 $\pm$ 181.1&445.7 $\pm$ 181.6&77.5&12.5\\
0.5&3&715.3 $\pm$ 227.9&631.2 $\pm$ 196.5&631.2 $\pm$ 196.5&630.3 $\pm$ 196.5&0.0&0.0\\
0.5&5&804.5 $\pm$ 225.7&748.9 $\pm$ 230.4&748.9 $\pm$ 230.4&747.4 $\pm$ 232.3&0.0&0.0\\
\bottomrule
\end{tabular}
\centering
\caption{Aggregated objective value, $\numops=15$}
\label{tb:exp/agg-obj-n=15}
\end{table}

\begin{table}[H]
\begin{tabular}{cccccccc}
\toprule
&&\multicolumn{4}{c}{Time [s], average $\pm$ std}\\
\cmidrule(l){3-6}
$\alpha_3$&$\devmax$&Greedy&Tabu&LBBD&BB\\
\midrule
0.1&0&0.0 $\pm$ 0.0&0.1 $\pm$ 0.0&0.1 $\pm$ 0.1&0.0 $\pm$ 0.0\\
0.1&3&0.0 $\pm$ 0.0&0.1 $\pm$ 0.0&0.3 $\pm$ 0.3&0.0 $\pm$ 0.0\\
0.1&5&0.0 $\pm$ 0.0&0.1 $\pm$ 0.0&0.7 $\pm$ 0.9&0.0 $\pm$ 0.0\\
0.3&0&0.0 $\pm$ 0.0&0.1 $\pm$ 0.0&0.1 $\pm$ 0.1&0.0 $\pm$ 0.0\\
0.3&3&0.0 $\pm$ 0.0&0.1 $\pm$ 0.0&0.5 $\pm$ 0.8&0.0 $\pm$ 0.0\\
0.3&5&0.0 $\pm$ 0.0&0.1 $\pm$ 0.0&1.1 $\pm$ 1.5&0.0 $\pm$ 0.0\\
0.5&0&0.0 $\pm$ 0.0&0.1 $\pm$ 0.0&0.2 $\pm$ 0.4&0.0 $\pm$ 0.0\\
0.5&3&0.0 $\pm$ 0.0&0.1 $\pm$ 0.0&1.0 $\pm$ 1.3&0.0 $\pm$ 0.0\\
0.5&5&0.0 $\pm$ 0.0&0.1 $\pm$ 0.0&2.0 $\pm$ 2.2&0.0 $\pm$ 0.0\\
\bottomrule
\end{tabular}
\centering
\caption{Aggregated running time, $\numops=5$}
\label{tb:exp/agg-run-n=5}
\end{table}

\begin{table}[H]
\begin{tabular}{cccccccc}
\toprule
&&\multicolumn{4}{c}{Time [s], average $\pm$ std}\\
\cmidrule(l){3-6}
$\alpha_3$&$\devmax$&Greedy&Tabu&LBBD&BB\\
\midrule
0.1&0&0.0 $\pm$ 0.0&0.1 $\pm$ 0.0&4.2 $\pm$ 7.2&1.0 $\pm$ 1.1\\
0.1&3&0.0 $\pm$ 0.0&0.1 $\pm$ 0.0&483.1 $\pm$ 536.0&1.3 $\pm$ 1.2\\
0.1&5&0.0 $\pm$ 0.0&0.2 $\pm$ 0.0&781.2 $\pm$ 502.6&1.6 $\pm$ 1.3\\
0.3&0&0.0 $\pm$ 0.0&0.1 $\pm$ 0.0&28.0 $\pm$ 65.9&1.6 $\pm$ 1.2\\
0.3&3&0.0 $\pm$ 0.0&0.1 $\pm$ 0.0&759.5 $\pm$ 525.0&2.0 $\pm$ 1.3\\
0.3&5&0.0 $\pm$ 0.0&0.2 $\pm$ 0.0&1036.6 $\pm$ 370.4&2.4 $\pm$ 1.4\\
0.5&0&0.0 $\pm$ 0.0&0.1 $\pm$ 0.0&18.8 $\pm$ 30.3&2.0 $\pm$ 1.7\\
0.5&3&0.0 $\pm$ 0.0&0.1 $\pm$ 0.0&961.7 $\pm$ 395.3&2.6 $\pm$ 1.8\\
0.5&5&0.0 $\pm$ 0.0&0.2 $\pm$ 0.0&1141.9 $\pm$ 208.5&3.0 $\pm$ 1.8\\
\bottomrule
\end{tabular}
\centering
\caption{Aggregated running time, $\numops=10$}
\label{tb:exp/agg-run-n=10}
\end{table}

\begin{table}[H]
\begin{tabular}{cccccccc}
\toprule
&&\multicolumn{4}{c}{Time [s], average $\pm$ std}\\
\cmidrule(l){3-6}
$\alpha_3$&$\devmax$&Greedy&Tabu&LBBD&BB\\
\midrule
0.1&0&0.0 $\pm$ 0.0&0.1 $\pm$ 0.0&137.8 $\pm$ 230.6&896.8 $\pm$ 479.5\\
0.1&3&0.0 $\pm$ 0.0&0.3 $\pm$ 0.1&1056.3 $\pm$ 379.3&1062.5 $\pm$ 361.3\\
0.1&5&0.0 $\pm$ 0.0&0.4 $\pm$ 0.1&1080.8 $\pm$ 354.8&1081.4 $\pm$ 320.2\\
0.3&0&0.0 $\pm$ 0.0&0.1 $\pm$ 0.0&423.0 $\pm$ 423.6&1103.6 $\pm$ 311.3\\
0.3&3&0.0 $\pm$ 0.0&0.2 $\pm$ 0.0&1169.1 $\pm$ 186.9&1153.8 $\pm$ 210.3\\
0.3&5&0.0 $\pm$ 0.0&0.3 $\pm$ 0.1&1174.4 $\pm$ 153.8&1170.6 $\pm$ 183.7\\
0.5&0&0.0 $\pm$ 0.0&0.1 $\pm$ 0.0&460.9 $\pm$ 442.3&1136.0 $\pm$ 192.3\\
0.5&3&0.0 $\pm$ 0.0&0.2 $\pm$ 0.0&1199.0 $\pm$ 0.0&1200.0 $\pm$ 0.0\\
0.5&5&0.0 $\pm$ 0.0&0.3 $\pm$ 0.1&1199.0 $\pm$ 0.0&1200.0 $\pm$ 0.0\\
\bottomrule
\end{tabular}
\centering
\caption{Aggregated running time, $\numops=15$}
\label{tb:exp/agg-run-n=15}
\end{table}

\subsection{Second experiment: evaluation of the tabu search on large instances}
\subsubsection{Instances}
The instances were generated in a similar manner as in Section~\ref{sec:exp/comparison} with the exception that the number of the operations was fixed to 100.
\subsubsection{Results}
We compared a simple Earliest Due Date First (EDF) ordering rule with \emph{Tabu}.
For this experiment, the stopping condition of \emph{Tabu} was modified: instead of running for the pre-specified number of iterations, \emph{Tabu} is stopped if the objective was not improved in the last 50 iterations; the rest of the parameters are the same as in Table~\ref{tab:heur/tabu/params}.

The average objective values with a standard deviation obtained by the algorithms are shown in Table~\ref{tb:exp/agg-obj-n=100} and the average running times of the algorithms with a standard deviation are shown in Table~\ref{tb:exp/agg-run-n=100}.
We can see that \emph{Tabu} is able to find significantly better solutions than the EDF rule within one minute.
On average, over all instances, solutions found by \emph{Greedy} and \emph{Tabu} algorithms are $26.5\%$ and $40.2\%$ better than EDF, respectively.
\emph{Tabu} improves the initial solution found by \emph{Greedy} algorithm by $18.6\%$.

\begin{table}[H]
\begin{tabular}{ccccc}
\toprule
&&\multicolumn{3}{c}{Objective value, average $\pm$ std}\\
\cmidrule(l){3-5}
$\alpha_3$&$\devmax$&EDF&Greedy&Tabu\\
\midrule
0.1&0&20101.8 $\pm$ 5958.5&12854.1 $\pm$ 4866.3&8898.8 $\pm$ 4639.8\\
0.1&3&32171.8 $\pm$ 6047.1&19130.7 $\pm$ 4704.1&15966.2 $\pm$ 4792.4\\
0.1&5&39465.8 $\pm$ 6345.3&25117.9 $\pm$ 5198.9&19330.0 $\pm$ 4874.8\\
0.3&0&22810.2 $\pm$ 6448.9&17310.3 $\pm$ 5870.9&12195.6 $\pm$ 5746.0\\
0.3&3&35228.0 $\pm$ 7264.9&23290.3 $\pm$ 5583.9&20129.2 $\pm$ 5949.5\\
0.3&5&41343.4 $\pm$ 6993.2&29707.7 $\pm$ 5998.8&24152.6 $\pm$ 5894.9\\
0.5&0&29926.3 $\pm$ 6233.8&25845.9 $\pm$ 6109.9&19967.6 $\pm$ 5976.9\\
0.5&3&41215.0 $\pm$ 6057.4&33252.8 $\pm$ 5604.8&29037.8 $\pm$ 6223.5\\
0.5&5&45246.8 $\pm$ 6183.4&39422.1 $\pm$ 6058.4&34274.7 $\pm$ 6862.9\\
\bottomrule
\end{tabular}
\centering
\caption{Aggregated objective value, $n=100$}
\label{tb:exp/agg-obj-n=100}
\end{table}

\begin{table}[H]
\begin{tabular}{ccccc}
\toprule
&&\multicolumn{3}{c}{Time [s], average $\pm$ std}\\
\cmidrule(l){3-5}
$\alpha_3$&$\devmax$&EDF&Greedy&Tabu\\
\midrule
0.1&0&0.0 $\pm$ 0.0&0.0 $\pm$ 0.0&3.0 $\pm$ 0.4\\
0.1&3&0.0 $\pm$ 0.0&0.0 $\pm$ 0.0&11.8 $\pm$ 3.9\\
0.1&5&0.0 $\pm$ 0.0&0.0 $\pm$ 0.0&28.9 $\pm$ 12.3\\
0.3&0&0.0 $\pm$ 0.0&0.0 $\pm$ 0.0&3.0 $\pm$ 0.4\\
0.3&3&0.0 $\pm$ 0.0&0.0 $\pm$ 0.0&11.2 $\pm$ 3.4\\
0.3&5&0.0 $\pm$ 0.0&0.0 $\pm$ 0.0&26.0 $\pm$ 8.1\\
0.5&0&0.0 $\pm$ 0.0&0.0 $\pm$ 0.0&3.0 $\pm$ 0.4\\
0.5&3&0.0 $\pm$ 0.0&0.0 $\pm$ 0.0&9.8 $\pm$ 2.5\\
0.5&5&0.0 $\pm$ 0.0&0.0 $\pm$ 0.0&16.2 $\pm$ 5.8\\
\bottomrule
\end{tabular}
\centering
\caption{Aggregated running time, $n=100$}
\label{tb:exp/agg-run-n=100}
\end{table}

\section{Conclusion}
\label{sec:conc}
In this work, we tackled the scheduling problem of satisfying the energy consumption limits by the manufacturing or production companies with high energy demand under production uncertainties.
The problem with uncertainty often occurs in reality since the carried-out schedule is usually different from the proposed baseline schedule.

Our main contribution is an algorithm that finds the optimal robust baseline schedule for the fixed permutation of the operations.
This algorithm can be used either as: (i) a method for making the existing baseline schedule robust or (ii) operator incorporated in a scheduling algorithm that seeks the optimal permutation.

We employed the algorithm in two exact (logic-based Benders decomposition and Branch-and-Bound) algorithms and one heuristic algorithm (tabu search).
The exact algorithms were evaluated on small instances.
The Branch-and-Bound algorithm is less sensitive to increase of the maximum deviation $\devmax$, whereas the logic-based Benders decomposition approach is viable if the number of cutting constraints needed to generate robust schedules is low (which is typical if the energy limits are not very tight and the maximum deviation is small).
In the experiments, we showed that with increasing the maximum deviation and the lower bound on the power consumption of the operations the instances are harder to solve.

To assess the scalability of our approach, the tabu search heuristic was evaluated on large instances with 100 operations.
On average, the instances were solved within 1 minute w.r.t.\ to the stopping criteria (the search stops if the objective value is not improved in the last 50 iterations).
On average, the tabu search improved the objective by $40.2\%$ over the Earliest Due Date First ordering rule.


\section*{Acknowledgement}
\label{sec:ack}
The work in this paper was supported by the Technology Agency of the Czech Republic under the Centre for Applied Cybernetics TE01020197.


\bibliography{/home/modosist/workspace/bibliography/bibfile}

\appendix
\clearpage
\section{Proofs}
\begin{lem}
\label{lem:app/basestarts-order-latestarts}
  Let $\basesched[1], \basesched[2]$ be two baseline schedules with the same permutation $\oppermfn$ such that for some position $\opposmax \in \intinterval{1}{\numops}$ holds $\forall \oppos \in \intinterval{1}{\opposmax}: \basestart[1]{\opperm{\oppos}} \le \basestart[2]{\opperm{\oppos}}$.
  Then $\forall \oppos \in \intinterval{1}{\opposmax}: \latestartfn{\basesched[1]}{\opperm{\oppos}} \le \latestartfn{\basesched[2]}{\opperm{\oppos}}$.
\end{lem}
\begin{proof}
  Proof by induction on $\oppos$
  \begin{enumerate}
    \item basis, $\oppos = 1$: $\latestartfn{\basesched[1]}{\opperm{1}} = \basestart[1]{\opperm{1}} + \devmax \le \basestart[2]{\opperm{1}} + \devmax = \latestartfn{\basesched[2]}{\opperm{1}}$
    \item induction step, $1 < \oppos \le \opposmax$:
      \begin{align}
        \latestartfn{\basesched[1]}{\opperm{\oppos}} &= \max(\basestart[1]{\opperm{\oppos}}, \latestartfn{\basesched[1]}{\opperm{\oppos - 1}} + \proctime{\opperm{\oppos - 1}}) + \devmax \\
        &\le \max(\basestart[2]{\opperm{\oppos}}, \latestartfn{\basesched[2]}{\opperm{\oppos - 1}} + \proctime{\opperm{\oppos - 1}}) + \devmax \\
        &= \latestartfn{\basesched[2]}{\opperm{\oppos}}
      \end{align}
  \end{enumerate}
\end{proof}

\begin{lem}
\label{lem:optsched/shiftdev}
  Let $\basesched$ be a baseline schedule and $\oppermfn$ be the corresponding permutation.
  Let $\shiftschedfnsym(\basesched, \opposmax, \mt)$ be a right-shift schedule for some $\opposmax \in \intinterval{1}{\numops}$ and $\mt \in \intinterval{\basestart{\opperm{\opposmax}}}{\latestartfn{\basesched}{\opperm{\opposmax}}}$.
  Then there exist scenario $\devsit \in \devsitsset$ such that \hbox{$\forall \oppos \in \intinterval{1}{\opposmax}: \realschedfn[\basesched][\devsit]_{\opperm{\oppos}} = \shiftschedfnsym(\basesched, \opposmax, \mt)_{\opperm{\oppos}}$}.
\end{lem}
\begin{proof}
  Let $\shiftsched = \shiftschedfnsym(\basesched, \opposmax, \mt)$ and $\realsched = \realschedfn[\basesched][\devsit]$.
  First we prove that \hbox{$\forall \oppos \in \intinterval{1}{\opposmax}: \shiftstart{\opperm{\oppos}} \in \intinterval{\basestart{\opperm{\oppos}}}{\latestartfn{\basesched}{\opperm{\oppos}}}$}.
  The property holds trivially from the definition for $\oppos = \opposmax$.
  To prove the property for $\oppos < \opposmax$ assume by contradiction that $\oppos < \opposmax$ is the largest position such that $\shiftstart{\opperm{\oppos}} \not\in \intinterval{\basestart{\opperm{\oppos}}}{\latestartfn{\basesched}{\opperm{\oppos}}}$.
  Since $\shiftstart{\opperm{\oppos}} \le \latestartfn{\basesched}{\opperm{\oppos}}$ holds from the definition of the right-shift start time, it must $\shiftstart{\opperm{\oppos}} < \basestart{\opperm{\oppos}} \le \latestartfn{\basesched}{\opperm{\oppos}}$ and therefore.
  \begin{align}
    \shiftstart{\opperm{\oppos}}=\min(\latestartfn{\basesched}{\opperm{\oppos}}, \shiftstart{\opperm{\oppos + 1}} - \proctime{\opperm{\oppos}}) =\shiftstart{\opperm{\oppos + 1}} - \proctime{\opperm{\oppos}}
  \end{align}
  However, this leads to contradiction
  \begin{equation}
    \shiftstart{\opperm{\oppos + 1}} - \proctime{\opperm{\oppos}} = \shiftstart{\opperm{\oppos}} < \basestart{\opperm{\oppos}} \le \basestart{\opperm{\oppos + 1}} -\proctime{\opperm{\oppos}}  \le \shiftstart{\opperm{\oppos + 1}}-\proctime{\opperm{\oppos}}
  \end{equation}

  Now we prove that if $\devsit$ is defined as
  \begin{align}
    \dev{\opperm{\oppos}} = 
    \begin{dcases*}
      \shiftstart{\opperm{1}} - \basestart{\opperm{1}} & $\oppos= 1$\\
      \shiftstart{\opperm{\oppos}} - \max(\basestart{\opperm{\oppos}}, \shiftstart{\opperm{\oppos - 1}} + \proctime{\opperm{\oppos - 1}}) & $\oppos\in\intinterval{2}{\opposmax}$
    \end{dcases*}
  \end{align}
  then for each position $\oppos \in \intinterval{1}{\opposmax}$ holds that $\dev{\opperm{\oppos}} \in \intinterval{0}{\devmax}$ and $\realstart{\opperm{\oppos}} = \shiftstart{\opperm{\oppos}}$.
  Proof by induction on $\oppos$
  \begin{enumerate}
    \item basis, $\oppos = 1$:
      \begin{itemize}
        \item $\dev{\opperm{1}} \in \intinterval{0}{\devmax}$:
          \begin{align}
            &\dev{\opperm{1}} = \shiftstart{\opperm{1}} - \basestart{\opperm{1}} \ge 0 \\
            &\dev{\opperm{1}} = \shiftstart{\opperm{1}} - \basestart{\opperm{1}} \le \latestartfn{\basesched}{\opperm{1}} - \basestart{\opperm{1}} = \basestart{\opperm{1}} + \devmax- \basestart{\opperm{1}} = \devmax
          \end{align}

        \item $\realstart{\opperm{1}} = \shiftstart{\opperm{1}}$: $\quad\realstart{\opperm{1}} = \basestart{\opperm{1}} + \dev{\opperm{1}} = \basestart{\opperm{1}} + \shiftstart{\opperm{1}} - \basestart{\opperm{1}} = \shiftstart{\opperm{1}}$
      \end{itemize}

    \item induction step, $1 < \oppos \le \opposmax$:
      \begin{itemize}
        \item $\dev{\opperm{\oppos}} \ge 0$: consider cases
          \begin{enumerate}
            \item $\basestart{\opperm{\oppos}}\ge\shiftstart{\opperm{\oppos - 1}} + \proctime{\opperm{\oppos - 1}}$:
              \begin{align}
                &\dev{\opperm{\oppos}} = \shiftstart{\opperm{\oppos}} - \max(\basestart{\opperm{\oppos}}, \shiftstart{\opperm{\oppos - 1}} + \proctime{\opperm{\oppos - 1}}) = \shiftstart{\opperm{\oppos}} - \basestart{\opperm{\oppos}} \ge 0
              \end{align}

            \item $\basestart{\opperm{\oppos}}<\shiftstart{\opperm{\oppos - 1}} + \proctime{\opperm{\oppos - 1}}$:
              \begin{align}
                \dev{\opperm{\oppos}} &= \shiftstart{\opperm{\oppos}} - \max(\basestart{\opperm{\oppos}}, \shiftstart{\opperm{\oppos - 1}} + \proctime{\opperm{\oppos - 1}}) \\
                &=(\shiftstart{\opperm{\oppos}} - \proctime{\opperm{\oppos - 1}}) - \shiftstart{\opperm{\oppos - 1}}\\
                &\ge\min(\latestartfn{\basesched}{\opperm{\oppos - 1}}, \shiftstart{\opperm{\oppos}} - \proctime{\opperm{\oppos - 1}}) - \shiftstart{\opperm{\oppos - 1}} \\
                &= \shiftstart{\opperm{\oppos - 1}} - \shiftstart{\opperm{\oppos - 1}} \\
                &=0
              \end{align}
          \end{enumerate}

        \item $\dev{\opperm{\oppos}} \le \devmax$: consider cases
          \begin{enumerate}
            \item $\shiftstart{\opperm{\oppos - 1}} = \latestartfn{\basesched}{\opperm{\oppos - 1}}$: since
              \begin{align}
                \latestartfn{\basesched}{\opperm{\oppos}} &= \max(\basestart{\opperm{\oppos}}, \latestartfn{\basesched}{\opperm{\oppos - 1}} + \proctime{\opperm{\oppos - 1}}) + \devmax \\
                &= \max(\basestart{\opperm{\oppos}}, \shiftstart{\opperm{\oppos - 1}} + \proctime{\opperm{\oppos - 1}}) + \devmax
              \end{align}
              therefore
              \begin{equation}
                \dev{\opperm{\oppos}} = \shiftstart{\opperm{\oppos}} - \max(\basestart{\opperm{\oppos}}, \shiftstart{\opperm{\oppos - 1}} + \proctime{\opperm{\oppos - 1}}) = \shiftstart{\opperm{\oppos}} - \latestartfn{\basesched}{\opperm{\oppos}} + \devmax \le \devmax
              \end{equation}

            \item $\shiftstart{\opperm{\oppos - 1}} = \shiftstart{\opperm{\oppos}} - \proctime{\opperm{\oppos - 1}}$:
              \begin{equation}
                \dev{\opperm{\oppos}} = \shiftstart{\opperm{\oppos}} - \max(\basestart{\opperm{\oppos}}, \shiftstart{\opperm{\oppos - 1}} + \proctime{\opperm{\oppos - 1}}) = \shiftstart{\opperm{\oppos}} - \max(\basestart{\opperm{\oppos}}, \shiftstart{\opperm{\oppos}}) \le 0 \le \devmax
              \end{equation}
          \end{enumerate}

        \item $\realstart{\opperm{\oppos}} = \shiftstart{\opperm{\oppos}}$:
          \begin{align}
            \realstart{\opperm{\oppos}} &= \max(\basestart{\opperm{\oppos}}, \realstart{\opperm{\oppos - 1}} + \proctime{\opperm{\oppos - 1}})+ \dev{\opperm{\oppos}} \\
            &=\max(\basestart{\opperm{\oppos}}, \shiftstart{\opperm{\oppos - 1}} + \proctime{\opperm{\oppos - 1}})+ \dev{\opperm{\oppos}} \\
            &=\max(\basestart{\opperm{\oppos}}, \shiftstart{\opperm{\oppos - 1}} + \proctime{\opperm{\oppos - 1}})+ \shiftstart{\opperm{\oppos}} - \max(\basestart{\opperm{\oppos}}, \shiftstart{\opperm{\oppos - 1}} + \proctime{\opperm{\oppos - 1}}) \\
            &= \shiftstart{\opperm{\oppos}}
          \end{align}

      \end{itemize}
  \end{enumerate}
\end{proof}

\begin{lem}
\label{lem:app/shifted-greater-realised}
  Let $\basesched[1], \basesched[2]$ be two baseline schedules with the same permutation $\oppermfn$ such that for some position $\opposmax \in \intinterval{1}{\numops}$
  \begin{align}
    &\basestart[1]{\opperm{\opposmax}} = \basestart[2]{\opperm{\opposmax}} \\
    &\basestart[1]{\opperm{\opposmax}} \le \basestart[2]{\opperm{\opposmax}} \,,\quad \oppos \in \intinterval{1}{\opposmax -1}
  \end{align}
  Then for every $\mt \in \intinterval{\basestart{\opperm{\opposmax}}}{\latestartfn{\basesched[1]}{\opperm{\opposmax}}}$ and every $\devsit \in \devsitsset$ such that $\mt = \realschedfn[\basesched[1]][\devsit]_{\opperm{\opposmax}}$ it holds that
  \begin{align}
    &\realschedfn[\basesched[1]][\devsit]_{\opperm{\opposmax}} = \shiftschedfnsym(\basesched[2], \opposmax, \mt)_{\opperm{\opposmax}} \\
    &\realschedfn[\basesched[1]][\devsit]_{\opperm{\oppos}} \le \shiftschedfnsym(\basesched[2], \opposmax, \mt)_{\opperm{\oppos}} \,,\quad \oppos \in \intinterval{1}{\opposmax -1}
  \end{align}
\end{lem}
\begin{proof}
  Proof by induction on $\oppos$
  \begin{enumerate}
    \item basis, $\oppos = \opposmax$: holds from the assumptions.
    \item induction step, $1 \le \oppos < \opposmax$: consider two cases
      \begin{enumerate}
        \item $\latestartfn{\basesched[2]}{\opperm{\oppos}} > \shiftschedfnsym(\basesched[2], \opposmax, \mt)_{\opperm{\oppos + 1}} - \proctime{\opperm{\oppos}}$: then
          \begin{equation}
            \realschedfn[\basesched[1]][\devsit]_{\opperm{\oppos}} \le \realschedfn[\basesched[1]][\devsit]_{\opperm{\oppos + 1}} -  \proctime{\opperm{\oppos}} \le \shiftschedfnsym(\basesched[2], \opposmax, \mt)_{\opperm{\oppos + 1}} - \proctime{\opperm{\oppos}} = \shiftschedfnsym(\basesched[2], \opposmax, \mt)_{\opperm{\oppos}}
          \end{equation}
        \item $\latestartfn{\basesched[2]}{\opperm{\oppos}} \le \shiftschedfnsym(\basesched[2], \opposmax, \mt)_{\opperm{\oppos + 1}} - \proctime{\opperm{\oppos}}$: then from Lemma~\ref{lem:app/basestarts-order-latestarts}
          \begin{equation}
            \realschedfn[\basesched[1]][\devsit]_{\opperm{\oppos}} \le \latestartfn{\basesched[1]}{\opperm{\oppos}} \le \latestartfn{\basesched[2]}{\opperm{\oppos}} = \shiftschedfnsym(\basesched[2], \opposmax, \mt)_{\opperm{\oppos}}
          \end{equation}
      \end{enumerate}
  \end{enumerate}
\end{proof}

\begin{lem}
\label{lem:app/starts-order-intersection}
  Let $\sched[1], \sched[2]$ be two schedules (not necessarily baseline, realised, etc.) with the same permutation $\oppermfn$ such that for some position $\opposmax \in \intinterval{1}{\numops}$
  \begin{align}
    &\start[1]{\opperm{\opposmax}} = \start[2]{\opperm{\opposmax}} \\
    &\start[1]{\opperm{\oppos}} \le \start[2]{\opperm{\oppos}} \,,\quad \oppos \in \intinterval{1}{\opposmax -1}
  \end{align}
  Let $\metint \in \metintsset$ be arbitrary metering interval such that $\metintproctimesym(\metint,\opperm{\opposmax},\start[1]{\opperm{\opposmax}}) > 0$.
  Then
  \begin{equation}
    \sum_{\oppos = 1}^{\opposmax} \metintproctimesym(\metint,\opperm{\oppos},\start[1]{\opperm{\oppos}}) \cdot \power{\opperm{\oppos}} \le \sum_{\oppos = 1}^{\opposmax} \metintproctimesym(\metint,\opperm{\oppos},\start[2]{\opperm{\oppos}}) \cdot \power{\opperm{\oppos}}
  \end{equation}
\end{lem}
\begin{proof}
  The Lemma obviously holds for metering intervals such that $\metintstart{\metint} \ge \start[1]{\opperm{\opposmax}}$, therefore assume that $\metintstart{\metint} < \start[1]{\opperm{\opposmax}} < \metintend{\metint}$.
  We prove the Lemma by showing
  \begin{equation}
    \forall \oppos \in \intinterval{1}{\opposmax}: \metintproctimesym(\metint,\opperm{\oppos},\start[1]{\opperm{\oppos}}) \le \metintproctimesym(\metint,\opperm{\oppos},\start[2]{\opperm{\oppos}})
  \end{equation}
  The inequality obviously holds for $\oppos = \opposmax$, therefore assume $\oppos < \opposmax$.
  Since \hbox{$\start[1]{\opperm{\oppos}} +\proctime{\opperm{\oppos}} \le \start[1]{\opperm{\opposmax}}$} and \hbox{$\start[2]{\opperm{\oppos}} +\proctime{\opperm{\oppos}} \le \start[1]{\opperm{\opposmax}}$}, it holds that
  \begin{align}
    \metintproctimesym(\metint,\opperm{\oppos},\start[1]{\opperm{\oppos}}) &= \max(0, \start[1]{\opperm{\oppos}} +\proctime{\opperm{\oppos}} - \max(\metintstart{\metint}, \start[1]{\opperm{\oppos}})) \\
    \metintproctimesym(\metint,\opperm{\oppos},\start[2]{\opperm{\oppos}}) &= \max(0, \start[2]{\opperm{\oppos}} +\proctime{\opperm{\oppos}} - \max(\metintstart{\metint}, \start[2]{\opperm{\oppos}}))
  \end{align}
  Now consider function $f(x) = x + \proctime{\opperm{\oppos}} - \max(\metintstart{\metint}, x)$.
  It is easy to see that $f(x)$ is non-decreasing in $x$ and since from assumption we know that \hbox{$\start[1]{\opperm{\oppos}} \le \start[2]{\opperm{\oppos}}$}, it holds that
  \begin{align}
    \metintproctimesym(\metint,\opperm{\oppos},\start[1]{\opperm{\oppos}}) = \max(0, f(\start[1]{\opperm{\oppos}})) \le \max(0, f(\start[2]{\opperm{\oppos}})) = \metintproctimesym(\metint,\opperm{\oppos},\start[2]{\opperm{\oppos}})
  \end{align}
\end{proof}

\begin{thm}
\label{lem:app/shifted-dominance}
  Let $\basesched$ be some baseline schedule with corresponding permutation $\oppermfn$.
  Let $\opposmax \in \intinterval{1}{\numops}$, $\realsched$ be some realised schedule of operations  $\opperm{1},\opperm{2},\dots,\opperm{\opposmax}$ and $\shiftsched = \shiftschedfnsym(\basesched, \opposmax, \realstart{\opperm{\opposmax}})$.
  Then
\begin{equation}
  \forall \metint \in \metintsset: \metintproctimesym(\metint,\opperm{\opposmax},\realstart{\opperm{\opposmax}}) > 0 \implies \sum_{\oppos=1}^{\opposmax} \metintproctimesym(\metint,\opperm{\oppos},\realstart{\opperm{\oppos}}) \cdot \power{\oppos} \le \sum_{\oppos=1}^{\opposmax} \metintproctimesym(\metint,\opperm{\oppos},\shiftstart{\opperm{\oppos}}) \cdot \power{\oppos}
\end{equation}
\end{thm}
\begin{proof}
  From Lemma~\ref{lem:app/shifted-greater-realised} it holds that
  \begin{align}
    &\realstart{\opperm{\opposmax}} = \shiftstart{\opperm{\opposmax}} \\
    &\realstart{\opperm{\oppos}} \le \shiftstart{\opperm{\oppos}} \,,\quad \oppos \in \intinterval{1}{\opposmax -1}
  \end{align}
  By applying Lemma~\ref{lem:app/starts-order-intersection} on $\sched[1]=\realsched$ and $\sched[2]=\shiftsched$, the Theorem is proven.
\end{proof}

\begin{thm}
\label{thm:app/earliest-schedule-is-robust-optimal}
  Let $\basesched$ be a baseline schedule with corresponding permutation $\oppermfn$.
  If every operation starts at its earliest robust time in $\basesched$ then $\basesched$ is robust and optimal for permutation $\oppermfn$.
\end{thm}
\begin{proof}
  $\,$
  \begin{enumerate}
    \item \emph{Robustness}: The robustness of $\basesched$ is follows from the definition of the robust baseline start time, see Eq.~\ref{eq:optsched/robust-start-time}.

    \item \emph{Optimality}: To illustrate the proof, we will use the following figure with different schedules
  \begin{figure}[H]
    \centering
    \includegraphics[scale=0.7]{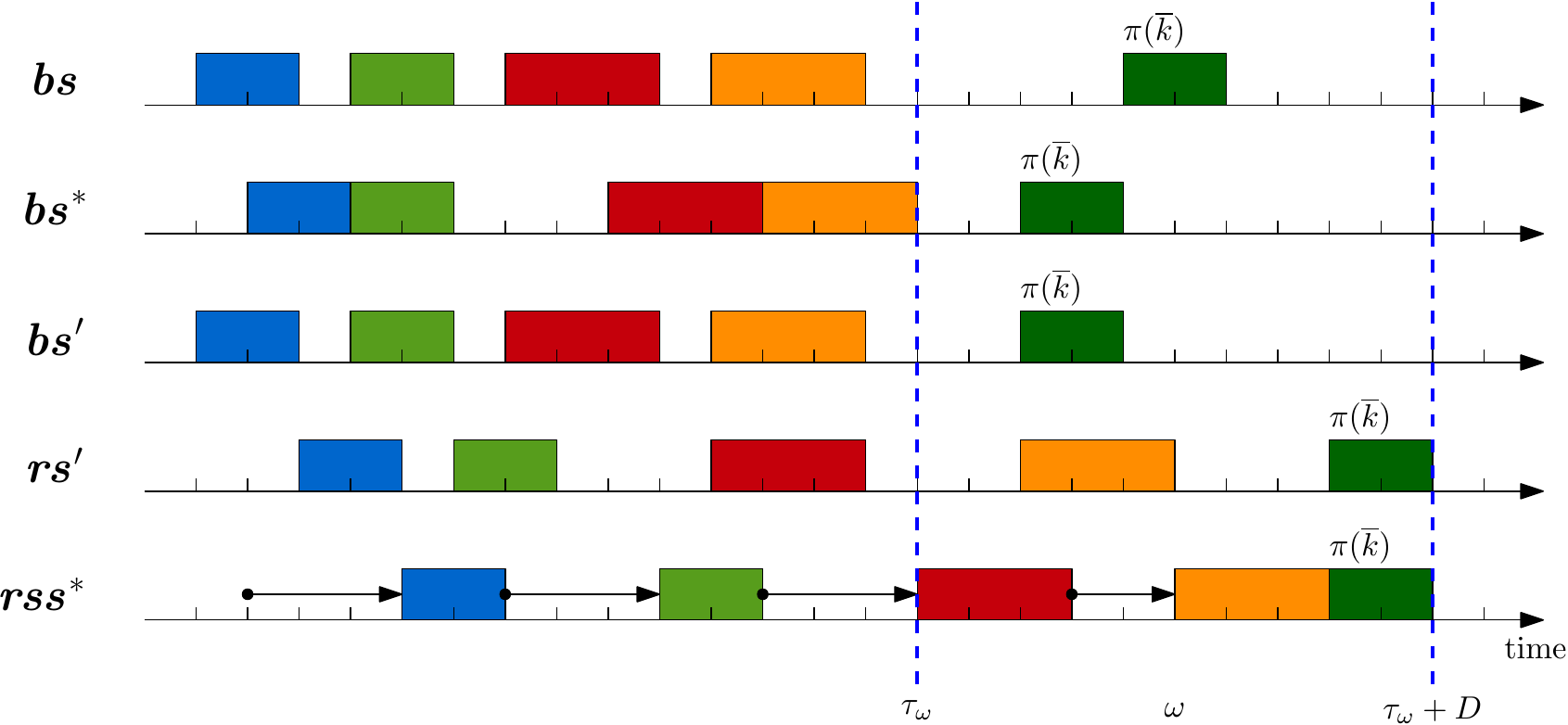}
    \caption{Example illustration of the schedules.}
\label{fig:app/optimal-schedules}
  \end{figure}

      Let $\baseschedoptmm$ be the optimal schedule for permutation $\oppermfn$.
      We need to show that \hbox{$\forall \oppos \in \intinterval{1}{\opposmax}: \basestart{\opperm{\oppos}} \le \basestartopt{\opperm{\oppos}}$}.
      Assume by contradiction that $\opposmax$ is the first position in $\oppermfn$ such that
      \begin{align}
        &\basestart{\opperm{\opposmax}} > \basestartopt{\opperm{\opposmax}} \\
        &\basestart{\opperm{\oppos}} \le \basestartopt{\opperm{\oppos}}\,,\quad \oppos \in \intinterval{1}{\opposmax-1}
      \end{align}
      Construct schedule $\basesched[\prime]$ such that
      \begin{align}
        &\basestart[\prime]{\opperm{\opposmax}} = \basestartopt{\opperm{\opposmax}} \\
        &\basestart[\prime]{\opperm{\oppos}} = \basestart{\opperm{\oppos}}\,,\quad \oppos \in \intinterval{1}{\opposmax-1}
      \end{align}
      i.e. $\basesched[\prime]$ is the same as $\basesched$ with exception of operation $\opperm{\opposmax}$ that starts at time $\basestartopt{\opperm{\opposmax}}$.
      Since $\basestart[\prime]{\opperm{\opposmax}}$ is not the earliest robust time in $\basesched$, there exist some realised schedule $\realsched[\prime]$ of $\basesched[\prime]$ such that
      \begin{equation}
        \maxenergy[\metint] < \sum_{\oppos = 1}^{\opposmax} \metintproctimesym(\metint,\opperm{\oppos},\realstart[\prime]{\opperm{\oppos}}) \cdot \power{\opperm{\oppos}}
      \end{equation}
      in some metering interval $\metint \in \metintsset$.
      Since operations $\opperm{1}, \dots, \opperm{\opposmax - 1}$ start at their earliest robust time in both schedules $\basesched$ and $\basesched[\prime]$, metering interval $\metint$ must have non-zero intersection with $\opperm{\opposmax}$ in $\realsched[\prime]$.
      Construct right-shift schedule \hbox{$\shiftsched[*] = \shiftschedfnsym(\basesched[*], \opposmax, \realstart[\prime]{\opperm{\opposmax}})$}; the construction is possible since from Lemma~\ref{lem:app/basestarts-order-latestarts} it holds that $\realstart[\prime]{\opperm{\opposmax}} \in \intinterval{\basestart[*]{\opperm{\opposmax}}}{\latestartfn{\basesched[*]}{\opperm{\opposmax}}}$.
      By applying Lemma~\ref{lem:app/shifted-greater-realised} for \hbox{$\basesched[1] = \basesched[\prime], \basesched[2] = \basesched[*]$} and Lemma~\ref{lem:app/starts-order-intersection} for \hbox{$\sched[1] = \realsched[\prime], \sched[2] = \shiftsched[*]$} we conclude that
      \begin{equation}
        \maxenergy[\metint] < \sum_{\oppos = 1}^{\opposmax} \metintproctimesym(\metint,\opperm{\oppos},\realstart[\prime]{\opperm{\oppos}}) \cdot \power{\opperm{\oppos}} \le \sum_{\oppos = 1}^{\opposmax} \metintproctimesym(\metint,\opperm{\oppos},\shiftstart[*]{\opperm{\oppos}}) \cdot \power{\opperm{\oppos}}
      \end{equation}
      which is a contradiction.
  \end{enumerate}
\end{proof}

\begin{lem}
\label{lem:app/earlier-violates-relative}
  Let $\basesched$ be a baseline schedule with the corresponding permutation $\oppermfn$.
Assume that the baseline start times of operations $\opperm{1}, \opperm{2}, \dots, \opperm{\opposmax-1}$ are robust for some $\opposmax \in \intinterval{1}{\numops}$.
  Moreover, assume that there exists some realised schedule $\realsched$ of $\basesched$ such that for some metering interval $\metint \in \metintsset$
  \begin{equation}
    \maxenergy[\metint] < \sum_{\oppos = 1}^{\opposmax} \metintproctimesym(\metint,\opperm{\oppos},\realstart{\opperm{\oppos}}) \cdot \power{\opperm{\oppos}}
  \end{equation}
  holds and $\metintproctimesym(\metint,\opperm{\opposmax-1},\realstart{\opperm{\opposmax-1}}) > 0$.
  Then all baseline start times $\intinterval{\basestart{\opperm{\opposmax-1}}+\proctime{\opperm{\opposmax-1}}}{\realstart{\opperm{\opposmax}}}$ of $\opperm{\opposmax}$ are not robust.
\end{lem}
\begin{proof}
  To illustrate the proof, we will use the following figure with different schedules
  \begin{figure}[H]
    \centering
    \includegraphics[scale=0.65]{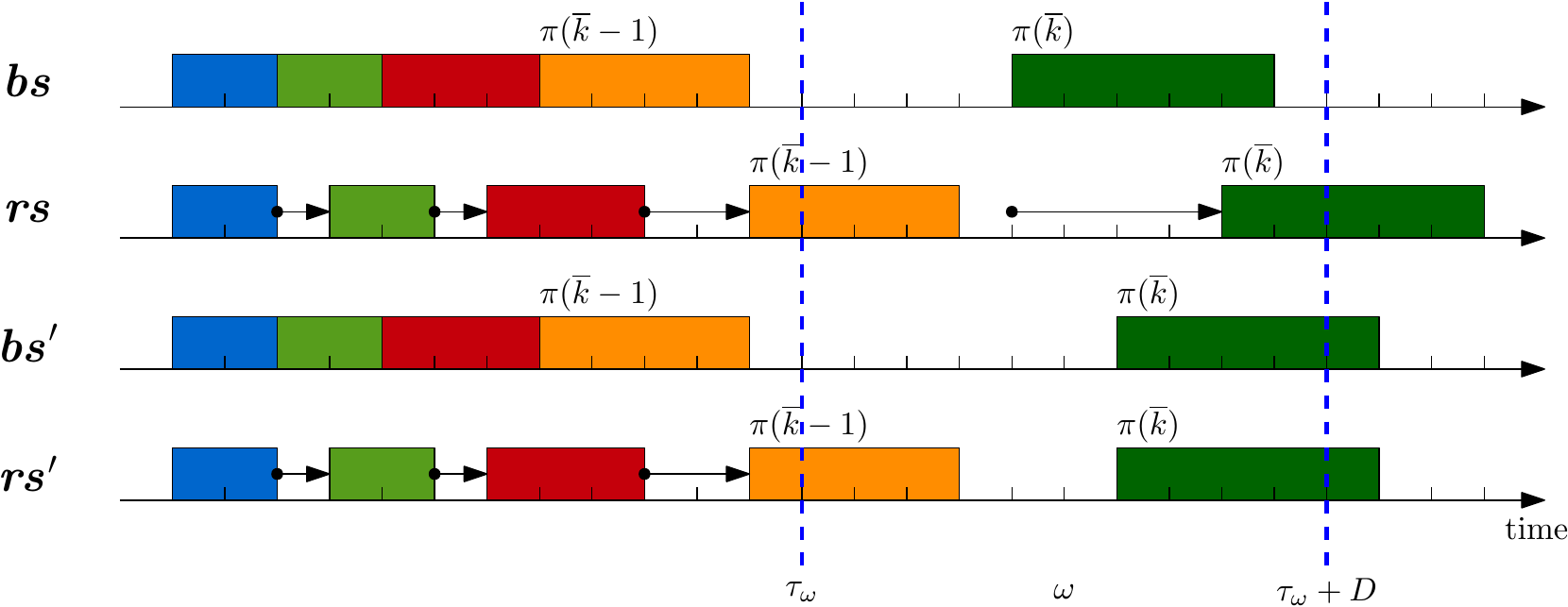}
    \caption{Example illustration of the schedules.}
\label{fig:app/earlier-violates}
  \end{figure}

  Assume by contradiction that there is baseline schedule $\basesched[\prime]$ such that
  \begin{align}
    &\basestart[\prime]{\opperm{\opposmax}} \in \intinterval{\basestart{\opperm{\opposmax-1}}+\proctime{\opperm{\opposmax-1}}}{\realstart{\opperm{\opposmax}}}\\
    &\basestart[\prime]{\opperm{\oppos}} = \basestart{\opperm{\oppos}} \,,\quad \oppos \in \intinterval{1}{\opposmax -1}
  \end{align}
  and $\basestart[\prime]{\opperm{\opposmax}}$ is robust.
  Construct schedule $\realsched[\prime]$ of $\basesched[\prime]$ such that
  \begin{align}
    &\realstart[\prime]{\opperm{\opposmax}} = \max(\basestart[\prime]{\opperm{\opposmax}}, \realstart[\prime]{\opperm{\opposmax-1}}+\proctime{\opperm{\opposmax-1}}) \\
    &\realstart[\prime]{\opperm{\oppos}} =  \realstart{\opperm{\oppos}}\,,\quad \oppos \in \intinterval{1}{\opposmax -1}
  \end{align}
  It is easy to see that $\realsched[\prime]$ is a realised schedule in which the deviation of $\opperm{\opposmax}$ is 0.

  Since $\realstart{\opperm{\opposmax}} \ge \realstart[\prime]{\opperm{\opposmax}} \ge \realstart[\prime]{\opperm{\opposmax-1}}+\proctime{\opperm{\opposmax-1}}$, it holds that \hbox{$\metintproctimesym(\metint,\opperm{\opposmax},\realstart{\opperm{\opposmax}}) \cdot \power{\opperm{\opposmax}}\le \metintproctimesym(\metint,\opperm{\opposmax},\realstart[\prime]{\opperm{\opposmax}})\cdot \power{\opperm{\opposmax}}$}; the argument is analogous to the one shown in proof of Lemma~\ref{lem:app/starts-order-intersection}.
  But this leads to contradiction
  \begin{equation}
    \maxenergy[\metint] < \sum_{\oppos = 1}^{\opposmax} \metintproctimesym(\metint,\opperm{\oppos},\realstart{\opperm{\oppos}}) \cdot \power{\opperm{\oppos}} \le \sum_{\oppos = 1}^{\opposmax} \metintproctimesym(\metint,\opperm{\oppos},\realstart[\prime]{\opperm{\oppos}})\cdot \power{\opperm{\oppos}}
  \end{equation}
\end{proof}


\end{document}